\documentclass[a4paper,12pt]{article}%

\usepackage[a4paper,top=2cm, bottom=2cm, left=2cm, right=2cm]{geometry}

\usepackage{amsmath}
\usepackage{amssymb}
\usepackage{graphicx}
\usepackage{geometry}
\usepackage{setspace}
\usepackage{placeins}
\usepackage{amsfonts}
\usepackage{subfig,graphicx,xcolor}
\usepackage{natbib}
\usepackage[colorinlistoftodos]{todonotes}
\usepackage{verbatim}
\usepackage{appendix}
\usepackage{tabularx}
\usepackage{dcolumn}
\usepackage{multirow}
\usepackage{booktabs}

\newcolumntype{d}[1]{D{.}{.}{#1}}

\doublespace

\begin{document}

\title{The public debt multiplier\thanks{This work acknowledges funding from the European
Union's Seventh Framework Programme under grant agreement n. 612796, for
the project MACFINROBODS (Integrated macrofinancial modelling for robust
policy design). We wish to thank Fabrice Collard, Martin Ellison, Jordi Gal\'{i} and Neil Rankin for comments. We also thank participants to the 6th edition of the Workshop in Macro Banking and Finance, to the 15th annual Dynare Conference and to the 3rd Annual NuCamp Conference.}}
\author{Alice Albonico\footnote{Corresponding author. Department of Economics, Management and Statistics, University of Milano - Bicocca,
Piazza dell'Ateneo Nuovo 1, 20126 Milan (Italy), Phone: +390264483041, E-mail: alice.albonico@unimib.it.}\\\textit{University of Milano - Bicocca}\\[-1pt]\textit{CefES}
\and Guido Ascari\footnote{Department of Economics,
University of Oxford, Manor Road, Oxford OX1 3UQ (United Kingdom), Phone: +441865271061, E-mail: guido.ascari@economics.ox.ac.uk.}\\\textit{University of Oxford}\\[-1pt]\textit{University of Pavia}
\and Alessandro Gobbi\footnote{Department of of Environmental Science and Policy, University of Milan,
Via Celoria 2, 20133 Milan (Italy), Phone: +390272342351, E-mail: alessandro.gobbi@unimi.it.}\\\textit{University of Milan}\\[-1pt]\textit{CefES}\\}
\maketitle

\begin{abstract} 
We study the effects on economic activity of a pure temporary change in government debt and the relationship between the debt multiplier and the level of debt in an overlapping generations framework. The debt multiplier is positive but quite small during normal times while it is much larger during crises. Moreover, it increases with the steady state level of debt. Hence, the call for fiscal consolidation during recessions seems ill-advised. Finally, a rise in the steady state debt-to-GDP level increases the steady state real interest rate providing more room for manoeuvre to monetary policy to fight deflationary shocks.

\vspace{0.4cm}


\end{abstract}




\thispagestyle{empty} 
\setcounter{page}{0}
\pagebreak

\section{Introduction}

In face of the Great Recession and the unprecedented fall in output, private consumption and investment spending, all advanced economies responded with a range of fiscal and monetary policy measures: increases in government spending, tax cuts, and various type of \textquotedblleft unconventional\textquotedblright monetary policy measures, given that monetary policy was unable to lower further the nominal interest rate already close or at the zero lower bound.

\begin{figure} [hb]
\begin{center}
	\subfloat[\footnotesize{Primary deficit to GDP ratio}\label{fig:figIntro1a}]{\includegraphics[scale=0.58]{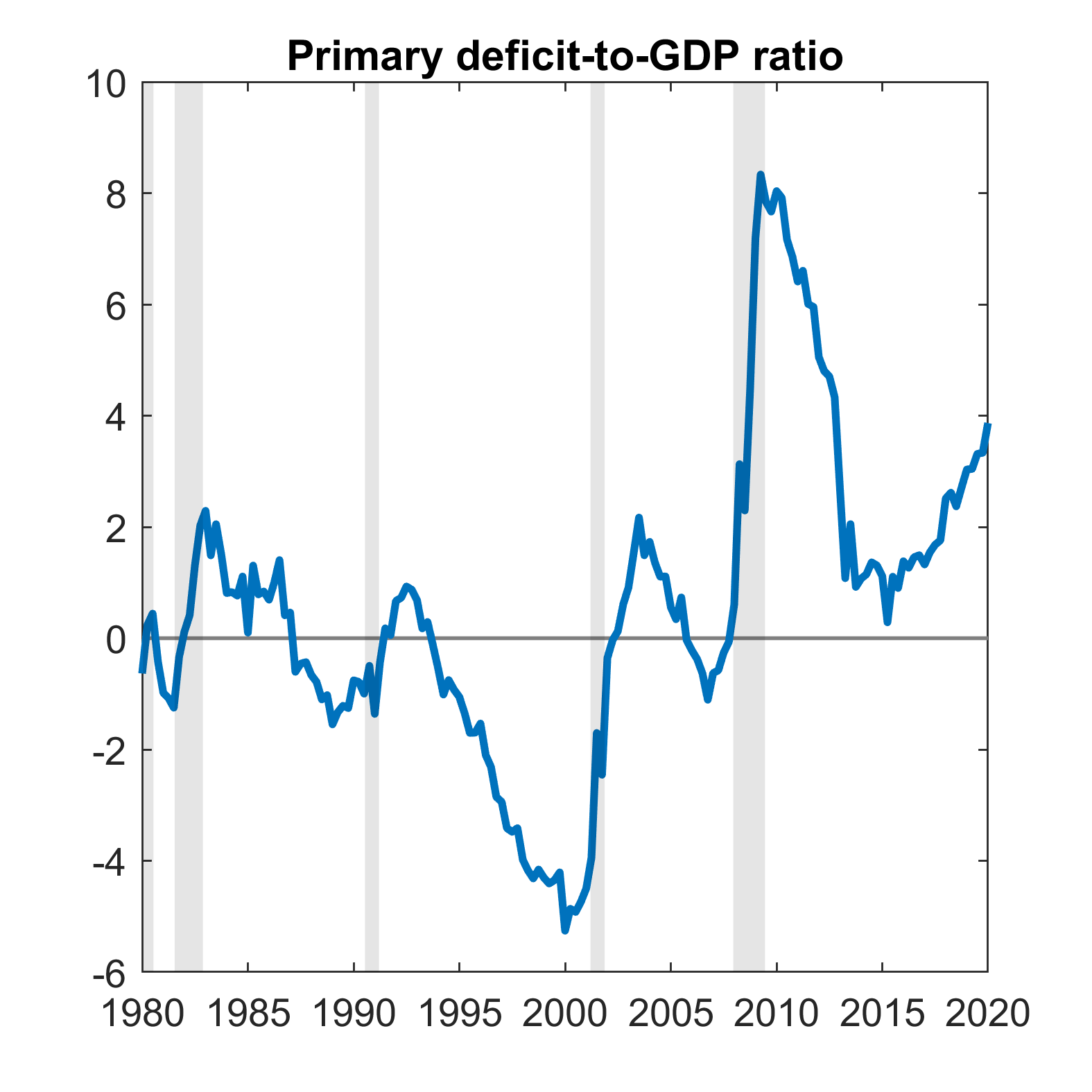}}
    \subfloat[\footnotesize{Debt-to-GDP ratio}\label{fig:figIntro1b}]{\includegraphics[scale=0.58]{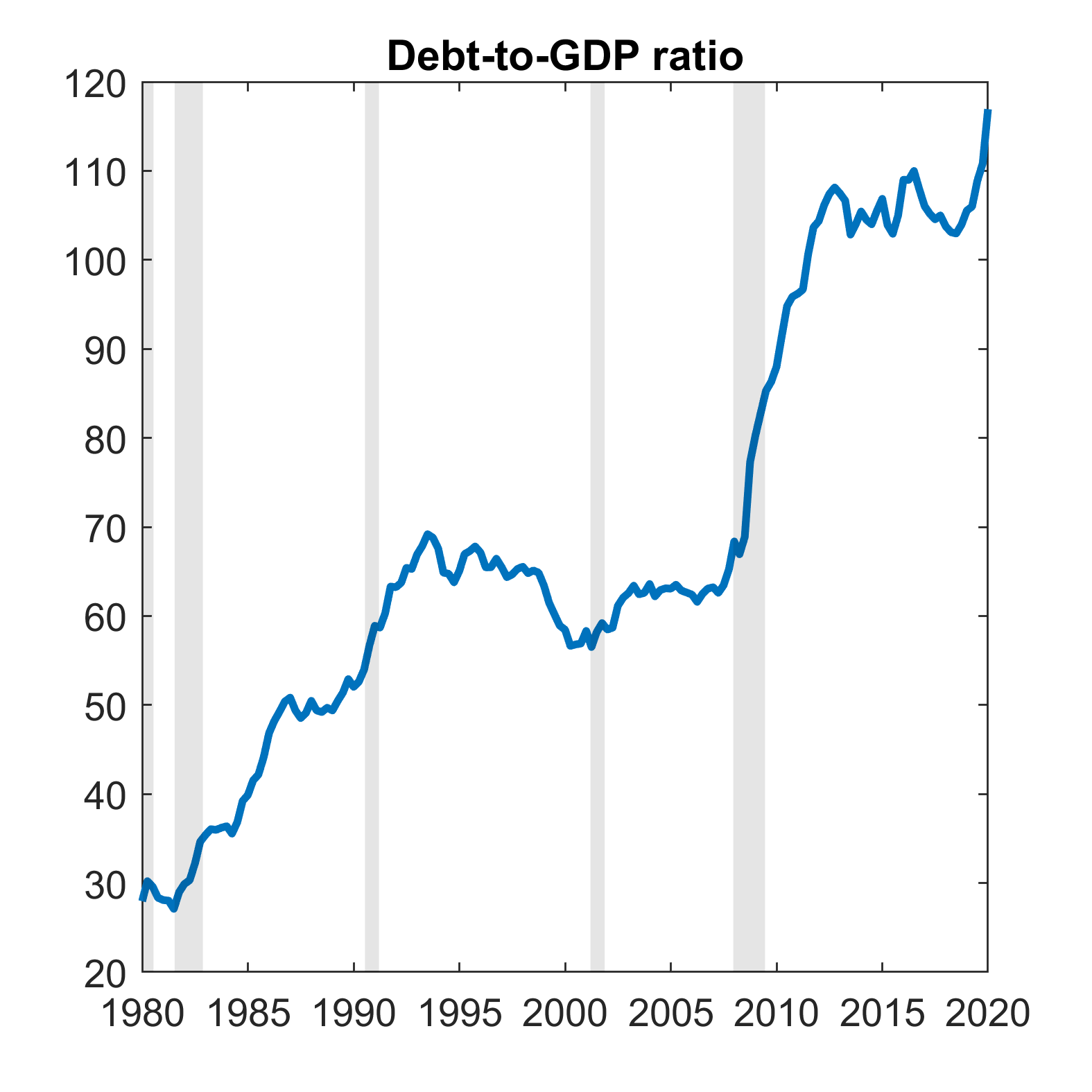}}
	\subfloat[\footnotesize{Inflation and federal funds rate}\label{fig:figIntro1c}]{\includegraphics[scale=0.58]{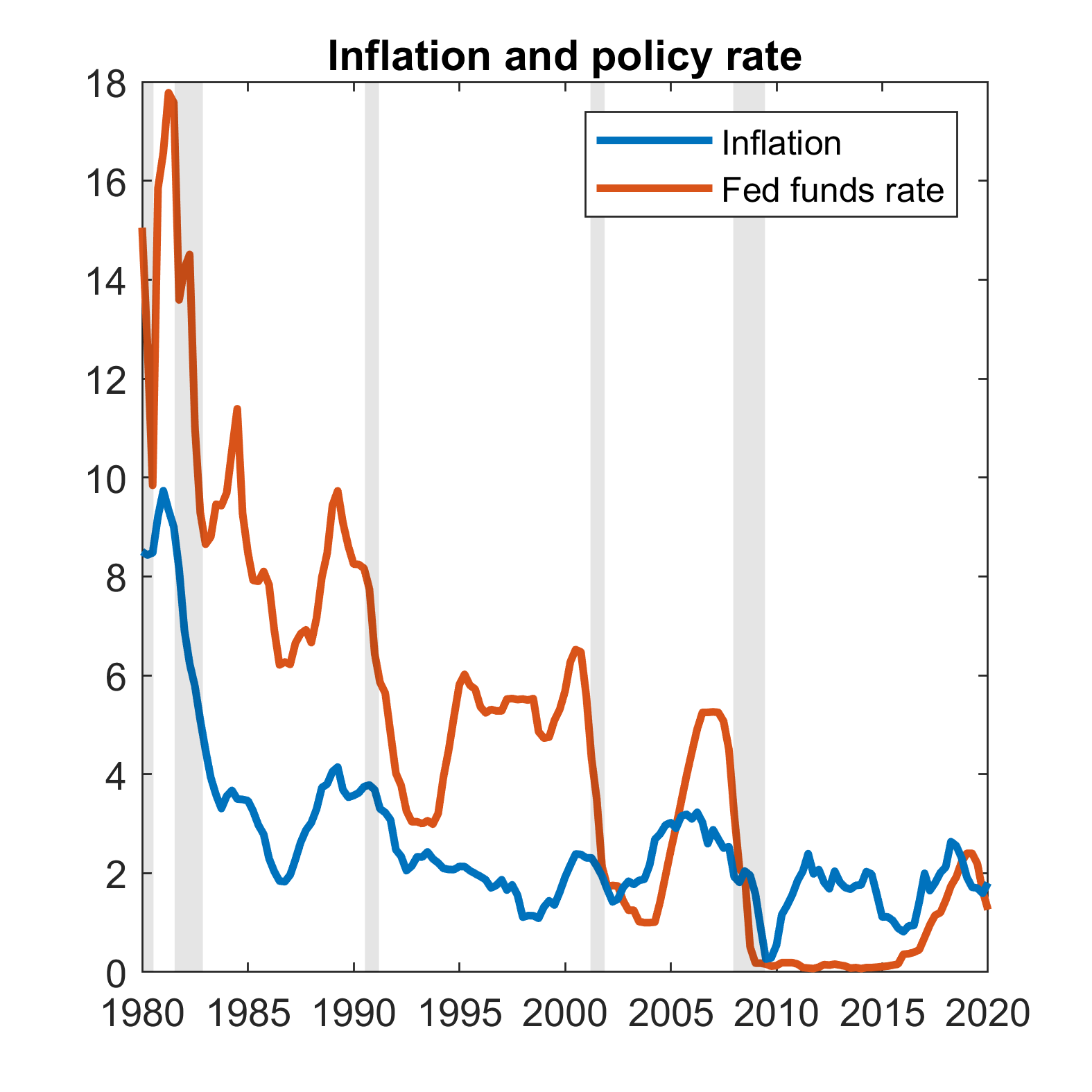}}
    \caption{The paths of the primary deficit to GDP ratio, debt-to-GDP ratio, inflation and policy rate in the U.S. Source: FRED database.\label{fig:figIntro1}}
\end{center}
\end{figure}
\FloatBarrier

As a result, the Great Recession deteriorated the fiscal positions of most advanced economies. 
Figures \ref{fig:figIntro1a} and \ref{fig:figIntro1b} show the expansion in both the fiscal deficit and the debt-to-GDP ratio for the United States starting in 2008. 
Monetary policy also reacted promptly decreasing sharply the nominal interest rate to a level very close to zero (see Figure \ref{fig:figIntro1c}). The level of inflation in the US rebounded after the initial drop and then started drifting down, remaining below target. Thus the US almost doubled the level of public debt from the pre-crisis level of about 60\% in 2007 to a level above 110\% in 2016. Other countries exhibit very similar dynamics, and the response to the crisis was clearly marked by an increase in public debt in most developed economies. In the euro area the increase in public debt triggered the sovereign debt crisis and a second recession. Evidently the government's aim was to support the collapsing level of aggregate demand. 

The increase in debt was brought about both by an increase in government spending and by a decrease in taxation. This boosted a large literature on  fiscal multiplier, both spending and tax multipliers.
Some works aimed at understanding the analytical mechanism of the fiscal multipliers, and its interaction with the zero lower bound \citep[see][]{cer11jpe,egge11nber,wood11aejm}. 
These works showed that multipliers for public spending are larger when the ZLB avoids the crowding out effect in private spending, thus amplifying the effect on output.
When nominal interest rates are positive and adjust according to a Taylor rule, the rise in inflation that follows an expansion in government spending causes the central bank to raise the nominal interest rate to counteract the initial increase in demand. Conversely, when the ZLB constraint is binding the nominal interest rate does not raise to curb inflationary pressures, thus not dampening demand.
Another strand of the literature focused on quantifying these multipliers empirically, using VARs, and theoretically, using DSGE models \citep[e.g.,][]{Leepetal10Jeco, ramey11JEL, zuba14ier, leepetal17AER,ramey19JEP}. 
Given the attention drawn by the Great Recession, many works investigated the possibility that fiscal multipliers vary depending on the state of the economy, and empirically investigates whether fiscal multipliers are higher during recessions than in normal times. The rational behind this assumption is that in a Keynesian world the economy will not always be characterized by full employment of resources, suggesting a higher multiplier during economic slack. The seminal paper by \citet{AuerGoro12} showed that fiscal multipliers are non-linear and state dependent. The papers by \citet{BlaLeigh13AER,BlaLeigh14IMF} and its preceding Box in the IMF World Economic Outlook 2012 have been widely discussed, and have clearly influenced the policy debate and possibly the policy stance in some countries. These results are corroborated by many other works \citep{AuerGoro13, caggianoetal15EJ, FMP15}. More recently, \citet{RamZub18} cast doubt on these results, because they estimate multipliers which are below one also during recession periods.

This very large literature investigates fiscal multipliers of both spending and different kinds of taxes. In this paper, we aim to fill a gap in this literature by instead analyzing the debt multiplier, that is, the output effect of a pure change in government debt, caused by the mere deferral of future taxes. 
This topic has acquired even greater importance in light of the COVID-19 epidemic.
In fact, while the Fed and other central banks reacted to the adverse shock by bringing interest rates to zero once again, the governments of many affected countries implemented a number of tax relief and tax deferral measures.  
Analysing the effects of tax cuts and the interplay between the level of debt and the stance of monetary policy is a key issue in the current economic debate.


What are in general the effects of a temporary and pure change in government debt on economic activity? It is well known in the literature that, assuming lump-sum taxes, a standard infinitely-lived representative (ILRA) agent model provides a very simple and disarming answer to this question: none. Ricardian equivalence would hold in such a model, so the effect of deferring lump-sum taxes would affect the economy only if (and to the extent that) taxes are distortionary. It is also well-known that in an overlapping generations (OLG) framework, the Ricardian equivalence does not hold because taxes levied at different times affect different cohorts, thus a pure deferral of taxes has a positive impact on private aggregate demand. As such, an increase in the level of public debt, caused by a cut in lump-sum taxes, would have a positive effect on output. This is the effect we want to study. One could consider the debt multiplier as a complement to the fiscal multipliers so far appeared in the literature, mainly derived in an ILRA framework. The debt multiplier would be the additional multiplier that one can get if any temporary fiscal policy measure would be financed through a pure temporary increase in public debt (i.e., a postponement of lump-sum taxes). The aim of this paper is to analyze this debt multiplier.

Hence, we move away from the infinitely-lived representative agent (ILRA) assumption and adopt an OLG setting. In particular, we investigate two questions: (i) How does the debt multiplier depend on the \textit{level} of public debt? (ii) Is the debt multiplier higher or lower in crisis time, that is, in a situation where there is a big negative demand shock so that the nominal interest rate hits the zero lower bound (ZLB)?

A clarification at this point is needed. What we call \textquotedblleft debt multiplier\textquotedblright is similar to what is often referred to in the literature as \textquotedblleft transfer multiplier\textquotedblright. \citet{GiaPe17} and \citet{Meh18} focus on the analysis of the transfer multiplier in two-agents models (respectively with a hand-to-mouth/Ricardians agents economy and with savers/borrowers consumers\footnote{Another example is \citet{BMP2013}.}). Notably, these are alternative ways of breaking the Ricardian equivalence. By using a different denomination, we would like to stress the role of debt as a fiscal instrument, and the idea that government debt is households' financial wealth, thus it is part of the supply of assets affecting the equilibrium real interest rate, as in \citet{ascaran13}.

Regarding question (i), in an ILRA framework the steady state real interest rate is pinned down by the subjective discount rate of the representative agent's utility function. Instead, in an OLG framework the real interest rate also depends on the amount of assets in the economy. The larger the amount of assets, the higher will be the steady state real interest rate needed for the OLG agents to be willing to hold those assets. Since government bonds represent net wealth in an OLG framework, an increase in the level of debt means an increase in the level of the real interest rate at steady state.\footnote{This relationship has already been analyzed in the literature, but mainly from an empirical point of view \citep[see for example,][for the US]{GalOrs05,enghub05,lau09}. The same result is shown in other OLG papers on fiscal policy, such as \citet{ascaran07, ascaran13, deve11}. However, these papers then assume a zero debt-to-GDP ratio in steady state.} Question (ii) arises naturally from the literature on fiscal multipliers, that shows that fiscal multipliers are larger when the nominal interest rate is stuck at the ZLB. 

This paper presents two lines of analysis. First, following \citet{egge11nber}, we provide analytical results about the sign of the debt multiplier by considering a linearized version of a simple OLG model.\footnote{The OLG environment has been used also by \citet{deve11} and \citet{smetra18} to deal with the non-trivial effect of the presence of government debt and the related policy issues. In particular, the first concentrates on the comparison of the effects of different fiscal interventions, the second deal with optimal monetary policy issues. We depart from these contributions by explicitly investigating the effects of a change in government debt per s\`{e} and the connection between fiscal interventions with the level of debt.}  Our findings indicate that the debt multiplier is positive even if the tax reduction is totally reversed in the future. We also find that debt multipliers increase with the level of debt, because of the positive relationship between fiscal multipliers and steady state real interest rates. Hence, the starting level of public debt affects the debt multiplier, especially so during a ZLB episode. In this respect, the calls for fiscal consolidation in recession times seem ill-advised. Second, to account for the presence of non-linear dynamics \citep[see,][]{lindetraba18}, we take the model in its original non-linear form and run a series of simulation exercises to quantify the debt multiplier. In normal times, or in a period of mild recession in which nominal rates remain positive, the multiplier is generally quite small, unless we shut off the wealth effect channel of government debt on labor supply. The debt multiplier is larger when the recession is severe and interest rates drop to zero. This reinstates the importance of fiscal stimulus when conventional monetary policy is impotent. 
Quantitatively, three elements are important for the size of the debt multiplier: the average life length (i.e., the survival probability), the income effect on labor supply and the persistence of the debt expansion. If we set the survival probability to 0.95, use GHH preferences, and consider a permanent expansion in debt, the debt multipliers in our model is equal to 0.74 on impact, not too distant from the estimates of the government spending multipliers reported by the recent literature. Finally, we also show that a high steady state level of debt could provide monetary policy with more room for manoeuvre in case of a deflationary shock, because it increases the steady state nominal rate, through a rise in the steady state real rate, for any given steady state inflation target.

The paper is organized as follows. Section \ref{sec:empirical_evidence}
presents some empirical evidence relating long-term real interest rates to the
level of debt-to-GDP. Section \ref{sec:model} describes the OLG model using two different specifications for the preferences of households. Section
\ref{sec:mult_analytical} contains the analytical derivation of debt
multipliers. Section \ref{sec:mult_numerical} investigates numerically the
size of multipliers and the relation with the level of debt. Section
\ref{sec:conclusions} concludes.

\section{Stylized facts \label{sec:empirical_evidence}}

One implication of the OLG framework is that the long-run debt-to-output ratio
determines the long-term real interest rate. The larger is the level of debt,
the higher the real interest rate. This is the key mechanism through which the
level of debt affects the debt multiplier. The question whether government debt affects the real interest rate has often been investigated in the literature. 
One possible channel through which this can occur is capital stock. If government spending crowds out the physical capital stock, then, from a simple production model, an exogenous increase in government debt causes the real interest rate to increase. 
Moreover, factors other than government debt can influence the determination of interest rates in credit markets \citep[see][]{enghub05}.

The insurgence of the sovereign debt crisis in 2010 has posed also
other interesting issues concerning the link between government debt and
interest rate spreads. Intuitively, as one country's government debt goes up,
the perception of investors on the risk of investing in that particular
country worsens, thus dampening demand for the country's government bonds.
This lowers the price of bonds and boosts risk premia and yields, which in
turn makes the debt burden heavier. Then, the country may need to increase
debt again to face higher interest payments, thus triggering a vicious cycle
of higher debt and higher interest rates.

\begin{figure}[ht]
\begin{center}
    \includegraphics[trim=50mm 90mm 50mm 90mm,clip, scale=0.8]{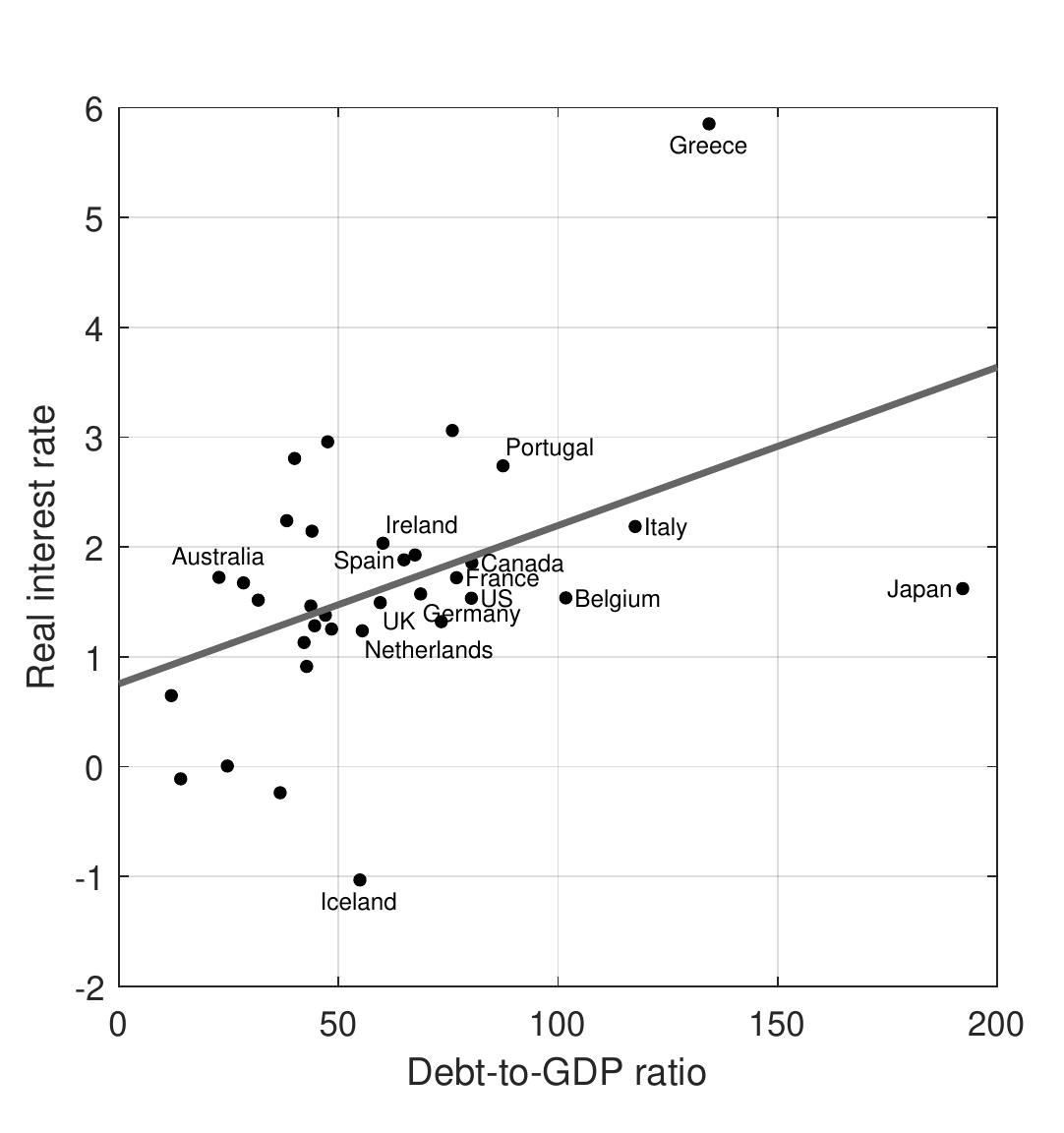}
    \caption{The relationship between debt-to-GDP ratio and
the real interest rate (2000-2018 averages).\label{fig:rr_scatter}}
\end{center}
\end{figure}

Figure \ref{fig:rr_scatter} shows the long-term real interest rate and the government debt-to-GDP
ratio for a set of countries. The long term interest rate is long-term
government bond yield (in most cases 10 year) adjusted for inflation
(calculated as the change in the GDP deflator). We use 2000-2018 averages to
indicate steady state values.\footnote{The countries considered here are: Australia, Austria, Belgium, Canada, Chile, Colombia, Czech Republic, Denmark, Finland, France, Germany, Greece, Hungary, Iceland, Ireland, Israel, Italy, Japan, Korea, Latvia, Lithuania, Luxembourg, Mexico, Netherlands, Norway, Poland, Portugal, Slovak Republic, Slovenia, Spain, Sweden, Switzerland, United Kingdom, United States. Data for general government debt as a percentage of GDP come from the Global Debt Database (International Monetary Fund). The long-term interest rates are downloaded from OECD Statistics (https://stats.oecd.org). Data for GDP deflator are retrieved from the World Development Indicators database (World Bank). We choose the 2000-2018 sample for data availability. However, we make robustness both on the set of countries and on the sample considered and the general result remains unchanged.} When average inflation is particularly high over
the sample we incur in negative values for the real interest rate. The scatter
plot captures the positive relationship between the two variables, as the
regression line is positively sloped. The case of Greece is evident, with a
high level of debt-to-GDP accompanied by a high real interest rate.

Thus, the
empirical evidence suggests that the long-term real interest rate is
positively related to the long-run level of debt as a percentage of GDP. In
the remainder of the paper, we will show that this relationship holds also
from a theoretical point of view.

\section{The model \label{sec:model}}

We use a dynamic general equilibrium model with overlapping generations
\textit{\`{a} la} \citet{bla95jpe}. Agents have an exogenous
probability, $q$ ($0<q<1$), of surviving to the next period. This framework
includes infinite lives as a special case, namely where $q=1$. Apart from the
characteristics concerning specifically overlapping generations we keep the
model as simple as we can and comparable to the standard infinitely lived
agents model of \citet{egge11nber}.

\subsection{Households}

The setup is very similar to \citet{ascaran13}.
Households maximize utility in consumption and leisure. They receive income
from working, from interests on bonds and from profits. At the same time, they
allocate their income stream between consumption and bond holdings, and they
pay lump-sum taxes.

As a benchmark case,
we consider the following log-log utility function:
\begin{equation} E_n
\sum_{t=n}^{\infty}\left(  \beta q\right)  ^{t-n}\xi_{t}\left[  \log\left(
C_{s,t}\right)  +\eta\log\left(  1-L_{s,t}\right)  \right]  \text{,}
\label{uf}%
\end{equation}
where $n$ is the current period and $s$ ($\leq n$) is the household's
birth-period. $C_{s,t}$ denotes consumption in period $t$ of a household born
in period $s$ and likewise for hours worked $L_{s,t}$. $\beta>0$ is the
discount factor, $\xi_{t}$ is a shock to preferences or to the discount factor
and $\eta>0$. In equilibrium a positive realization of $\frac{\xi_{t+1}}%
{\xi_{t}}$ will induce a rise in the effective discount factor so that
households want to save more. This will trigger a fall in consumption today
and lead to an economic recession which could imply zero nominal interest rates.

The optimization problem of the households consists of maximizing the utility
function subject to the budget constraint:%

\begin{equation}
P_{t}C_{s,t}+B_{s,t}^{N}=(1/q)\left(  1+i_{t-1}\right)  B_{s,t-1}^{N}%
+W_{t}L_{s,t}+P_{t}\left(D_{t}-T_{t}\right)\text{.}%
\end{equation}
$B_{s,t}^{N}$ are bond holdings of a household born in period $s$ in period
$t$. $P_{t}$, $W_{t}$ and $i_{t}$ indicate the price index, wage and the
nominal interest rate, respectively. $D_{t}$ and $T_{t}$ denote profits from
firms and lump-sum taxes. As in \citet{bla95jpe}, the households receive an
``annuity" at the gross rate $1/q$ on their financial wealth if they survive,
otherwise this wealth passes to the insurance company. The profits of the
insurance companies are zero in equilibrium.

From the first order conditions we obtain the following in the benchmark case:%
\begin{align}
\frac{W_{t}}{P_{t}}  &  =\eta\frac{C_{s,t}}{1-L_{s,t}}\text{,}%
\label{labor supply}\\
 C_{s,t} &  = E_{t}\left \{\frac{C_{s,t+1}}{\beta\left(  1+r_{t}\right)}  \frac{\xi_{t}}{\xi_{t+1}} \right \}\text{,} \label{euler equation}%
\end{align}
where $\left(1+r_{t}\right)  =\left(  1+i_{t}\right)  P_{t}/E_{t}P_{t+1}$ is the
real interest rate. Equation (\ref{labor supply}) is the individual labor supply, while equation (\ref{euler equation}) is the individual Euler equation. 

In this case the utility function implies the existence of a wealth effect between hours and consumption. Although the benchmark specification of the utility function is standard in OLG models, it implies that as agents age they will hit the constraint of non-negative labor supply, because $C_{s,t}$ increases with the
agent's age, $t-s$.\footnote{For a detailed discussion about this issue, see \citet{ascaran07}.} Hence, we consider also the case where this effect is absent, through the use of \citet{GHH88} (henceforth, GHH) preferences. Moreover, the wealth effect on labor supply has strong effects on the fiscal multipliers in DSGE models, so that the literature often employs similar type of preferences \citep[e.g.,][]{monapero08}. In the GHH case we have:

\begin{equation} E_n
\sum_{t=n}^{\infty}\left(  \beta q\right)  ^{t-n}\xi_{t}\log\left[
C_{s,t}-\left(  \eta/\varepsilon\right)  L_{s,t}^{\varepsilon}\right]
\text{,} \label{us GHH}%
\end{equation}
where $\varepsilon>1$. GHH preferences enables to avoid this problem as (\ref{labor supply}) is replaced by the following:
\begin{equation}
\frac{W_{t}}{P_{t}}=\eta L_{s,t}^{\varepsilon-1}\text{,} \label{ls GHH}%
\end{equation}
where there is no $C_{s,t}$ and thus the labor supply is independent of $s$.
At the same time (\ref{euler equation}) is replaced by:%
\begin{equation}
 C_{s,t}-\left(  \eta/\varepsilon\right)  L_{s,t}%
^{\varepsilon}  = E_{t}\left\{ \frac{C_{s,t+1}-\left(  \eta/\varepsilon\right)  L_{s,t+1}^{\varepsilon}}{\beta \left(  1+r_{t}\right)}  \frac{\xi_{t}}%
{\xi_{t+1}}\right\} . \label{ee GHH}%
\end{equation}

\paragraph{Aggregation.} 
We can derive the aggregate counterpart of equations
(\ref{labor supply}) and (\ref{euler equation}) by multiplying by $\left(
1-q\right)  q^{t-s}$ (individuals born in period $s$ and still alive in $t$)
and summing both sides for $s=-\infty,...,t$. The relationship between a
generic aggregate variable $X_{t}$ and its constituent individual variables
is:%
\begin{equation}
X_{t}=\sum_{s=t}^{-\infty}\left(  1-q\right)  q^{t-s}X_{s,t}\text{.}%
\end{equation}
After some algebra we obtain the following aggregate expressions:%
\begin{align}
\frac{W_{t}}{P_{t}}\left(  1-L_{t}\right)   &  =\eta C_{t}\text{,}%
\label{aggr ls}\\
E_{t}C_{t+1}+\frac{\left(  1-q\right)  }{\left(1+\eta\right)q}\frac{V_{t}}{E_{t}\delta_{t+1}}  &
=\beta E_{t} \left \{\left(  1+r_{t}\right)  \frac{\xi_{t+1}}%
{\xi_{t}} \right \}C_{t},
\label{aggr ee}%
\end{align}
where $\delta_{t}=1+q\beta E_{t} \left \{\frac{\xi_{t+1}}{\xi_{t}}\delta_{t+1}\right \}$.\footnote{This representation ignores complications due to Jensen’s inequality. It is, however, very useful to give an intuitive explanation of the effects of aggregation. Moreover, the error/approximation has no consequences for the results presented below. The analytical results in Section \ref{sec:mult_analytical} are derived from a first-order approximation, so the solution of the aggregate model is exact at this order. The numerical results in Section \ref{sec:mult_numerical} assume perfect foresight (or an MIT shock, since they derive from impulse responses following a debt shock). See also \citet{deve11} and \citet{smetra18}.} 
Note that for $q<1$, government debt represents net wealth and thereby affects consumption
spending. Following \citet{ascaran13}, we define financial wealth as:
$V_{s,t}=\frac{1}{q}\left(  1+i_{t}\right)  B_{s,t}^{N}$. Note that the
relationship of aggregate to individual financial wealth is slightly different
from the general one, being rather: $\frac{1}{q}V_{t}=\sum_{s=t}^{-\infty
}\left(  1-q\right)  q^{t-s}V_{s,t}$, because we include the annuity payout in
our definition of $V_{s,t}$. The aggregate Euler equation says that the growth
rate of aggregate consumption depends positively on the real interest rate, on
the growth rate of the shock and negatively on aggregate financial wealth.

For the model with GHH preferences (\ref{aggr ls}) and (\ref{aggr ee}) become:%
\begin{align}
\frac{W_{t}}{P_{t}}&=\eta L_{t}^{\varepsilon-1}, \label{aggr ls GHH}\\
E_{t}C_{t+1}+\frac{\left(  1-q\right)  }{q}\frac{V_{t}}{E_{t}\delta_{t+1}%
}-\left(  \eta/\varepsilon\right)  E_{t}L_{t+1}^{\varepsilon}%
&=\beta E_{t} \left \{\left(  1+r_{t}\right)  \frac{E_{t}\xi_{t+1}}{\xi_{t}}\right \}\left[
C_{t}-\left(  \eta/\varepsilon\right)  L_{t}^{\varepsilon}\right].  \label{aggr ee GHH}
\end{align}

Finally, the representative households decides how to allocate its consumption
expenditures among the different varieties of goods, indexed by $i\in\left[
0,1\right]  $. The consumption index is represented by a standard CES function:%
\begin{equation}
C_{s,t}=\left[  \int\limits_{0}^{1}C_{i,s,t}^{\frac{\theta-1}{\theta}%
}di\right]  ^{\frac{\theta}{\theta-1}}, \label{CES}%
\end{equation}
with $\theta>1$. The household maximizes (\ref{CES}) subject to the budget
constraint $\int\limits_{0}^{1}P\left(  i\right)  C\left(  i\right)
_{s,t}di=Z_{s,t}$, where $Z_{s,t}$ is total expenditure for goods. This leads
to the familiar constant elasticity demand function for good type $i$:%
\begin{equation}
C_{i,s,t}=\left(  \frac{P_{i,t}}{P_{t}}\right)  ^{-\theta}\frac{Z_{s,t}}%
{P_{t}}=\left(  \frac{P_{i,t}}{P_{t}}\right)  ^{-\theta}C_{s,t},
\label{good demand}%
\end{equation}
where%
\begin{equation}
P_{t}=\left[  \int\limits_{0}^{1}P_{i,t}^{1-\theta}di\right]  ^{\frac
{1}{1-\theta}}.
\end{equation}

\subsection{Firms}

The problem of firms is independent of households' preferences. There is a
continuum of firms indexed by $i\in\left[  0,1\right]  $, each producing a
differentiated good. All firms operate under monopolistic competition and use
the same technology represented by a Cobb-Douglas production function with aggregate labor as unique factor of production, so we have that  $Y_{i,t}=L_{i,t}^{\sigma}$ where $Y_{i,t}$ and $L_{i,t}$ are the output and the amount of labor employed by firm $i$.
Firms face the same demand schedule and take $P_{t}$, $W_{t}$ and $Y_{t}$ as given.

Prices are sticky \textit{\`{a} la} \citet{Calvo1983}, so that each firm may reset its price only with a probability $1-\alpha$ in any given period. With
probability $\alpha$ it keeps its price unchanged. The optimization problem of
a firm $i$ that adjusts its price $P_{i,n}$ in period $n$ is then:%
\begin{equation}
\max_{P_{i,n}}E_{n}\sum_{t=n}^{\infty}\alpha^{t-n}\Delta_{n,t}\frac
{P_{i,n}Y_{i,t}-W_{t}L_{i,t}}{P_{t}}%
\end{equation}
subject to the demand schedule
\begin{equation}
Y_{i,t}=\left(  \frac{P_{i,t}}{P_{t}}\right)  ^{-\theta}Y_{t}\text{,}%
\end{equation}
where the discount factor $\Delta_{n,t}$ is given by 
\begin{equation}
\Delta_{n,t}\equiv\left(  1+r_{n}\right)  ^{-1}\left(  1+r_{n+1}\right)
^{-1}...\left(  1+r_{t-1}\right)  ^{-1}\text{, \ \ \ }\Delta_{n,n}\equiv1.
\end{equation}

If we define $P^*_{n}$ as the new price set in period $n$ by firm $i$, solving
the optimization problem yields the following result:%
\begin{equation}
P^*_{n}=E_{n}\left[  \frac{\theta}{\theta-1}\frac{1}{\sigma}\frac{\sum_{t=n}^{\infty
}\alpha^{t-n}\Delta_{n,t}W_{t}Y_{t}^{\frac{1}{\sigma}}P_{t}^{\frac{\theta
}{\sigma}-1}}{\sum_{t=n}^{\infty}\alpha^{t-n}\Delta_{n,t}Y_{t}P_{t}^{\theta
-1}}\right]  ^{\frac{1}{1+\frac{\theta}{\sigma}-\theta}}\text{.} \label{Xn}%
\end{equation}
This expression says that the new price depends on current and expected future
values of aggregate output, the general price level and the wage level. All
firms resetting their prices in period $n$ will choose the same new price.

\subsection{Government}

The government's budget constraint in nominal terms is:%
\begin{equation}
P_{t}\left(  G_{t}-T_{t}\right)  +i_{t-1}B_{t-1}^{N}=\left(  B_{t}^{N}%
-B_{t-1}^{N}\right)  \text{,} \label{GOVbc}%
\end{equation}
where $G_{t}$ is public spending. In real terms we obtain:%
\begin{equation}
G_{t}-T_{t}=B_{t}-(1+r_{t-1})B_{t-1}\text{,} \label{realGOVbc}%
\end{equation}
where we defined $B_{t}=B_{t}^{N}/P_{t}$. Then, in line with \citet{ascaran13}, we express the government budget constraint in terms of the real government debt inclusive of interest payments,\footnote{Government debt should therefore be thought of as ‘indexed’ debt. More precisely, $B_{t}^{\prime}$ is the number of ‘real treasury bills’ issued, i.e. it is a promise to deliver $B_{t}^{\prime}$ units of the composite consumption good to the holders of the bonds at the start of period $t+1$.} 
which we denote as
$B_{t}^{\prime}=(1+r_{t})B_{t}$: 
\begin{equation}
G_{t}-T_{t}=(1+r_{t})^{-1}B_{t}^{\prime}-B_{t-1}^{\prime}\text{.}
\label{adjGOVbc}%
\end{equation}
Clearly only two of the three policy instrument can be chosen independently.
In what follows we will consider exogenous fiscal shocks to government debt,
thus leaving lump-sum tax $T_{t}$ to be determined implicitly by
(\ref{adjGOVbc}) as residual.

The central bank follows a standard contemporaneous Taylor rule if the nominal interest rates is positive. Otherwise it is constrained by the zero lower bound (see equation \eqref{TR}). 

\subsubsection{Market clearing}
Equilibrium in the goods market simply requires that:%
\begin{equation}
Y_{t}=C_{t}+G_{t}\text{.}%
\end{equation}

\section{Analytical investigation of debt
multipliers\label{sec:mult_analytical}}

In this section, we present the reduced form solution and we explain the
solution method to derive the multipliers for government debt. First, we
derive the multipliers in a positive interest rates environment, where the
monetary authority follows a Taylor rule. Second, we present the multipliers
in a zero interest rates context.

\subsection{Model solution}

\subsubsection{Benchmark case} \label{benchmark}

We derive a simplified version of the model after log linearizing the model so
that a generic variable $X_{t}$ can be approximated by $x_{t}=\frac{X_{t}%
-X}{X}\simeq\log\left(  \frac{X_{t}}{X}\right)  $.\footnote{Note that
$g_{t}=\frac{G_{t}-G}{Y}=\frac{G_{t}}{Y}$ as the steady state of $G_{t}$ is
assumed to be zero. We also define $i_{t}=\log\left(  1+i_{t}\right)  $, so
that $i_{t}$ is not in log-deviations. Where a lower case letter does not
exist we use the symbol ``$\; \widehat{}\; $" to indicate log deviations.}%
The following equations are the dynamic AD curve \eqref{AD}, the New Keynesian Phillips Curve \eqref{NKPC}, the Central bank rule for nominal interest rates \eqref{TR} and the dynamics of the auxiliary variable $\delta_t$ \eqref{delta}:
\begin{align}
E_{t}y_{t+1}  &  =\beta(1+\bar{r})y_{t}+\beta(1+\bar{r})\left(  i_{t}-E_{t}%
\pi_{t+1}\right)  +E_{t}g_{t+1}-\beta(1+\bar{r})g_{t}\label{AD}\\
&  -\beta(1+\bar{r})r_{t}^{e}-\left[  \beta(1+\bar{r})-1\right]  b_{t}%
^{\prime}+\left[  \beta\left(  1+\bar{r}\right)  -1\right]  E_{t}\hat{\delta}%
_{t+1}\text{,}\nonumber\\
\pi_{t}  &  =\left(  1+\bar{r}\right)  ^{-1}E_{t}\pi_{t+1}+\kappa\left[
y_{t}-\left(  1-L\right)  \sigma g_{t}\right]  \text{,}\label{NKPC}\\
i_{t}  &  =\max(0,r_{t}^{e}+\phi_{\pi}\pi_{t}+\phi_{y}y_{t})\text{,}%
\label{TR}\\
\hat{\delta}_{t}  &  =q\beta\hat{\delta}_{t+1}-q\beta\left[  r_{t}^{e}%
-\log\left(  1+\bar{r}\right)  \right]  , \label{delta}%
\end{align}
where we define $r_{t}^{e}=\log\left(  1+\bar{r}\right)  -\Delta\hat{\xi
}_{t+1}$, following \citet{egge11nber}, and
\[
\kappa=(1-\alpha)\left[  \alpha^{-1}-\left(  1+\bar{r}\right)  ^{-1}\right]
\left(  1-\theta+\frac{\theta}{\sigma}\right)  ^{-1}\frac{1}{1-L}\frac
{1}{\sigma}.
\]
Parameters $\phi_{\pi}$ and $\phi_{y}$ are the responses of the nominal interest rate to inflation and output, respectively. Moreover, $\bar{r}$ is the steady state value of the real interest rate, which
is derived endogenously and corresponds to:
\begin{equation}
\bar{r}=\frac{q\left(  1+\eta\right)}{\beta q\left(  1+\eta\right)-\left(  1-q\right)  \left(  1-\beta q\right)
\frac{B}{Y}}-1\text{.} \label{ssrbar}
\end{equation}
This expression states that the steady state real interest rate, and consequently the equilibrium solution, depends on the probability of surviving and on the steady state debt to output ratio. (\ref{ssrbar}) implies that a larger steady state debt-to-GDP level causes a higher level of the steady state real interest rate, which is consistent with the empirical evidence showed in Section \ref{sec:empirical_evidence}. The reason is the following. Agents are born with zero financial wealth, so they must accumulate it during their lifetime. In each period, thus, aggregate wealth drags on consumption growth in the aggregate Euler equation (\ref{aggr ee}) in the \cite{bla95jpe} OLG framework, because the agents that die in each period consume more that the one that are born (some works in the literature refer to this effect as ``generational turnover effect"). This has two important implications. First, the real interest rate, that keeps \textit{aggregate} consumption in steady state constant, is larger than the one that would keep the \textit{individual} consumption constant in steady state. According to (\ref{euler equation}), the latter is the same as in the ILRA model and equal to $(1/\beta-1)$.
Hence, differently from ILRA models, the following holds:
\begin{equation}
\beta\left(  1+\bar{r}\right)  >1\text{.}
\label{rlargerbeta}
\end{equation}
Given (\ref{euler equation}), condition (\ref{rlargerbeta}) implies that agents save and accumulate wealth during their lifetime, exactly because the real interest rate is larger than the subjective discount rate.

Second, the steady state real interest rate, $\bar{r}$, depends on financial wealth, i.e., on the supply of assets in the economy. The larger the amount of assets, the stronger the generational turnover effects and the larger should be the real interest rate to keep aggregate consumption constant. In simpler words, the supply and demand of assets determine the steady state real interest rate in this OLG framework, so that the larger the supply of assets in the economy the higher should be the real interest rate to make agents willing to hold that amount of assets.


This relationship has important implications for the model solution and we will come back to it in what follows. For the moment, it is important to stress that the steady state value of the debt affects the steady state real interest rate, and hence it affects the dynamics of the economy and, as we will show, the multipliers.

We consider temporary shocks. In computing the fiscal multipliers, we follow
the same approach in \citet{egge11nber} to which thus our results are
immediately comparable. 
To do so, we make the following assumptions. 
We define the long-run as the time at which the shock $r_{t}^{e}$ is at steady state.
The short-run is the period in which the economy is subject to temporary
disturbances and it is defined by $r_{t}^{e}=r_{S}^{e}$. This shock reverts
back to steady state with probability $(1-\mu)$. A negative
realization of $r_{t}^{e}$ implies that households want to save more so that
the real interest rate must decline to keep output constant. In general it can
be interpreted as any exogenous reason for a decline in spending. \citet{egge11nber} suggests it can be thought for example as an exogenous increase in the
probability of default by borrowers, which characterizes a banking crisis. As
in \citet{egge11nber}, we assume that fiscal policy is perfectly correlated
with the shock, that is, we consider government debt increases/decreases which
are a direct reaction to the shock, such that $b_{t}^{\prime}=b_{S}^{\prime}$
in the short run and $b_{t}^{\prime}=0$ in the long run (meaning that in the
long-run $B_{t}^{\prime}$ is back to the steady state).\footnote{Note that the
long run solution implies that $\pi_{t}$, $Y_{t}$, $\delta_{t}$ and $i_{t}$
remain at their steady state so that they are zero in log-deviations.} Here we
also implicitly assume that, for a given change in government debt, government
spending remains constant at its steady state and all the adjustment occurs
through a change in taxation.

Expectations about a generic variable $x_{t}$ (in deviation from its
steady state) are formed as follows:%
\begin{equation}
E_{t}x_{t+1}=\mu E^{S}_{t}x_{t+1}^{S}+\left(  1-\mu\right)  0=\mu E^{S}_{t}x_{t+1}^{S}, \label{exp}%
\end{equation}
where $S$ stands for ``short run". The notation $E^{S}_{t}$ is used as the expectation of the variable $x_{t+1}$ conditional on the shock being in the short run state. Accordingly, $x_{t+1}^{S}$ is the value of $x_{t+1}$ conditional
on the short run state. When the system is back to the steady state
the variable is equal to its steady state value, so that the deviation from
the steady state is zero. This applies to all the variables in the model.

The model can be solved through standard methods, such as the method of
undetermined coefficients. The solution of the model enables to determine an
algebraic expression for the multipliers of government spending and debt.

\subsection{Positive interest rates \label{sec:analitic_multi_noZLB}}

In this section we analyze a debt increase when interest rates are positive.
In this case, a standard Taylor rule describes monetary policy: a negative
realization of the shock $r_{t}^{e}$ prompts the reaction of the central bank
that cuts nominal rates to counteract the fall in output and inflation.
However, nominal rates remain positive as the shock $r_{t}^{e}$ is too small
to activate the ZLB constraint.

\begin{figure}[h!]
\begin{center}
    \includegraphics[trim=20mm 105mm 20mm 105mm,clip, scale=0.95]{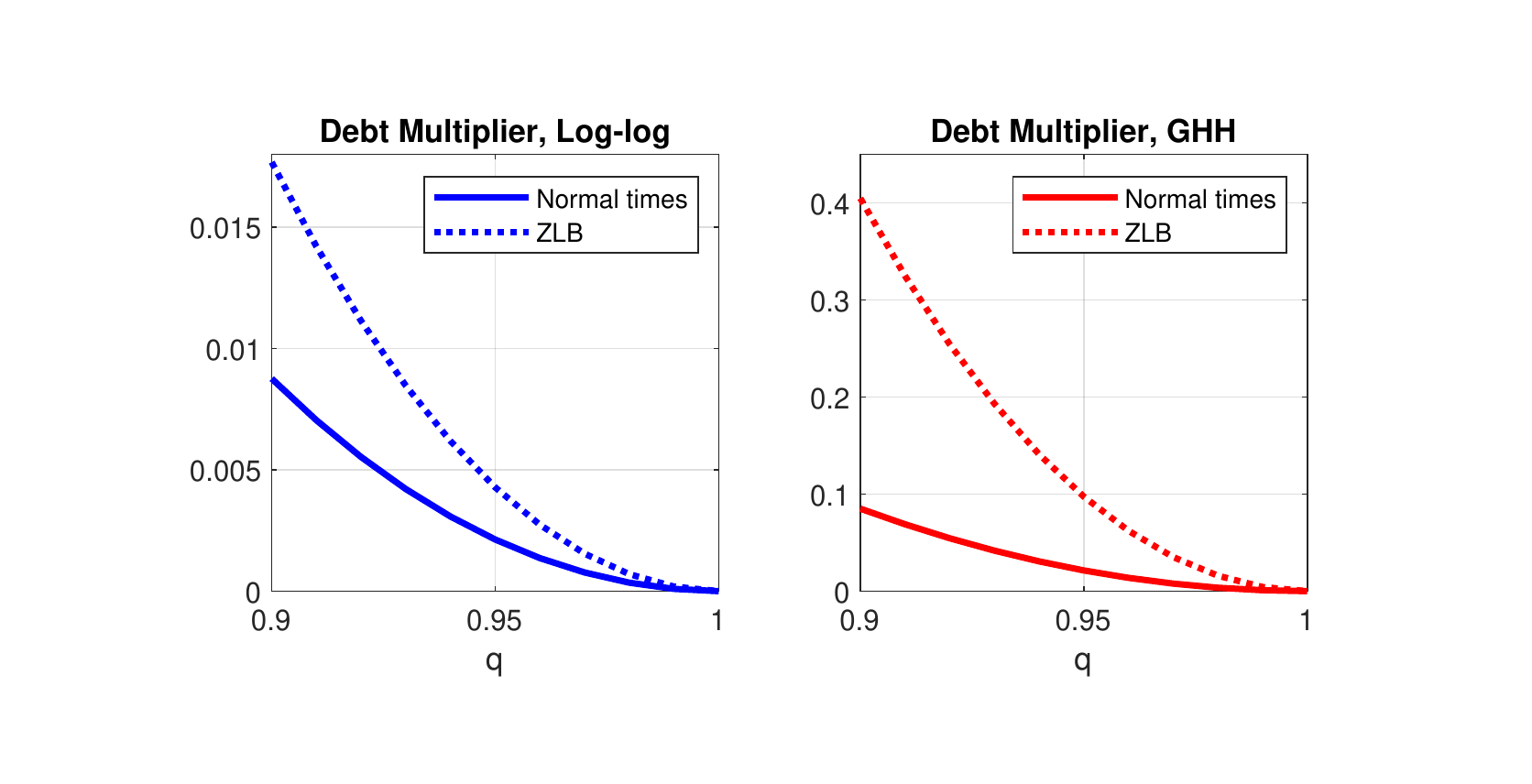}
    \caption{The debt multiplier for different values of the probability of survival. \newline {\protect\footnotesize \textit{Notes:} Analytic multipliers for the baseline calibration, see Table \ref{tab:calibration}.} }
    \label{fig:multiq}
\end{center}
\end{figure}

\begin{figure}[h!]
\begin{center}
    \includegraphics[trim=20mm 105mm 20mm 105mm,clip, scale=0.95]{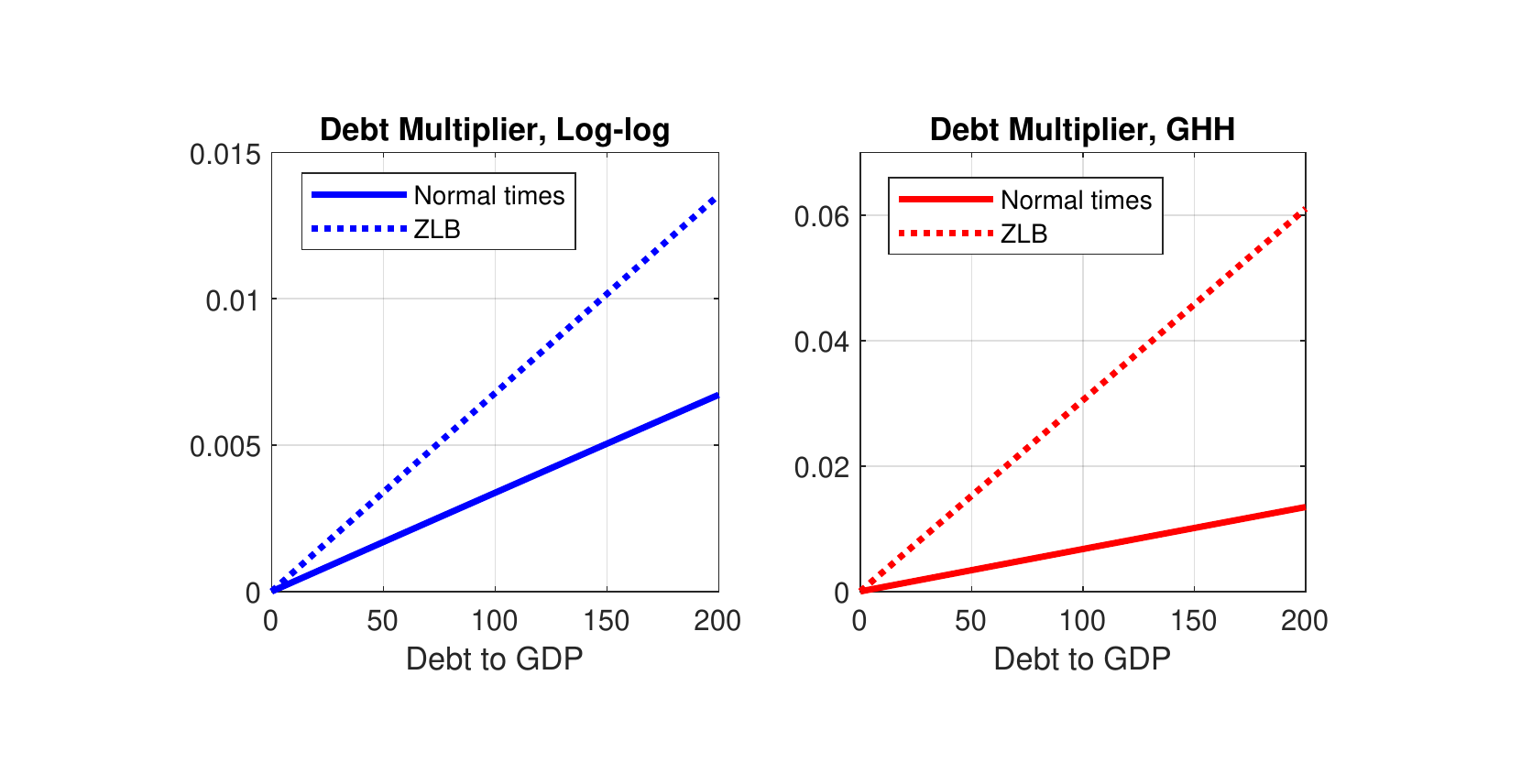}
    \caption{The debt multiplier for different values of the debt-to-GDP ratio. \newline {\protect\footnotesize \textit{Notes:} Analytic multipliers for the baseline calibration.} }
    \label{fig:multi2}
\end{center}
\end{figure}

We assume that government spending remains constant at its steady state. In
the short run taxes will adjust (decrease) in order to meet the government
budget constraint, implying a positive (higher than the steady state) level of
debt. Given the way expectations are formed, when the shock is switched off,
it is implicitly assumed that debt has to go back to its initial steady state
level. Otherwise, the steady state will be different from the initial one, and
that is not consistent with (\ref{exp}). Again the adjustment occurs through an increase in taxation. The impact government debt multiplier is:\footnote{The multiplier is computed as $dy^{S}/db^{\prime}_{S}$, where both $y^{S}$ and $b^{\prime}_{S}$ are in deviations from their steady state. Thus, a $1\%$ increase in debt leads to a $BM\%$ increase in output.}
\begin{equation}
BM=\frac{\left[  \beta(1+\bar{r})-1\right]  \left[  1-(1+\bar{r}%
)^{-1}\mu\right]  }{\left[  \beta(1+\bar{r})\left(  1+\phi_{y}\right)
-\mu\right]  \left[  1-(1+\bar{r})^{-1}\mu\right]  +\beta(1+\bar{r})\kappa\left(
\phi_{\pi}-\mu\right)  }. \label{bm olg}%
\end{equation}
Note that, given that $0<\mu<1$ and $\beta(1+\bar{r})>1$, the effect of an
increase in government debt is always positive if $\phi_{\pi}-\mu>0$, which is
likely to occur in standard models where the Taylor principle ($\phi_{\pi}>1$)
holds. In the case of a tax reduction, the increase in debt shifts out aggregate demand, stimulating production. Private consumption actually increases because in an OLG framework agents will experience a positive wealth effect, as they could be dead when taxes will adjusts upwards. The rise in consumption thus implies a higher demand and, in turn, higher output, because of nominal rigidities. Labor demand shifts upwards, because those firms which cannot adjust prices due to nominal rigidities produce more to cope with the increase in demand. 
While the positive wealth effect stimulates household consumption, it also reduces labour supply. Despite the shift in labor supply, the net effect is however positive for output. Note that in case of ILRA, private consumption will instead not respond because Ricardian equivalence holds. Indeed, the solid blue line in Figure \ref{fig:multiq} shows that the debt multiplier depends on the probability of surviving $q$ and it is equal zero when $q=1$.\footnote{In Figures \ref{fig:multiq} and \ref{fig:multi2} the probability $\mu$ of being in the short run state is set to 0.4. Remaining parameters are calibrated according to Table \ref{tab:calibration}.}


However, the multiplier is lower than one, that is, the effect of debt on output is less than proportional, since $\mu<1$ and $\phi_{\pi}-\mu>0$. Moreover, the multiplier is lower when the responses of monetary policy to both output ($\phi_{y}$) and inflation ($\phi_{\pi}$) are stronger. This is very intuitive, because the increase in debt will cause, \textit{ceteris
paribus,} an increase both in output and inflation, and then the more monetary policy responds to them, the lower will be the debt multiplier. The monetary policy response following a Taylor rule is partially crowding out the expansionary effects of an increase in debt, and the more so the more hawkish is monetary policy.

Last, but not least, the debt multiplier depends on $\bar{r}$, and hence it depends on the value of debt. This result has been neglected in the literature, and it arises naturally in an OLG framework, and it holds for any fiscal multiplier, not only the debt one. The solid blue line in Figure \ref{fig:multi2} shows that the debt multiplier increases with the level of debt. The coefficient on $b'_{t}$ on the RHS of (\ref{AD}), which derives from (\ref{aggr ee}), is an increasing function of $\bar{r}$, so that the effect of a change in $b'_{t}$ increases with the level of debt-to-gdp ratio. The intuition of this channel comes form the fact that in an OLG framework agents are saving during their lifetimes. Indeed, this coefficient in (\ref{AD}) is positive exactly because condition (\ref{rlargerbeta}) holds, as explained earlier. A temporary increase in the level of debt determines a temporary increase in the real interest rate. According to (\ref{euler equation}), agents are saving because the steady state real interest rate is larger then the subjective discount rate. However, a given marginal increase in the real interest rate would determine a lower response of savings (and a higher response of current consumption), the higher the starting (steady state) level of real interest rate, because the marginal utility is decreasing. In other words, the higher the steady real interest rate, the higher the rate of growth of individual consumption and hence the distance between the marginal utility between two consecutive periods. Hence, for a given temporary change in the real interest rate, agents will allocate the positive wealth effect due to an increase in the debt level relatively more to current consumption than to savings.

The multiplier of government debt is zero for the ILRA model under positive interest rates. In fact, in this case, $(1+\bar{r}=\beta)$, Ricardian equivalence holds and government debt plays no role in stimulating demand.

\subsubsection{GHH\ preferences}

We do the same exercise with the model characterized by GHH\ preferences.
Equations \eqref{AD} and \eqref{NKPC} become:%
\begin{align}
\frac{1}{\theta}E_{t}y_{t+1}  &  =\beta(1+\bar{r}_{GHH})\frac{1}{\theta}y_{t}%
+\frac{\beta(1+\bar{r}_{GHH})}{\zeta}\left(  i_{t}-E_{t}\pi_{t+1}\right)
\label{AD GHH}\\
&  + E_{t}g_{t+1}-\beta(1+\bar{r}_{GHH})g_{t}-\frac{\left[  \beta(1+\bar{r}%
_{GHH})-1\right]  }{\zeta}b_{t}^{\prime}\nonumber\\
&  -\frac{\beta(1+\bar{r}_{GHH})}{\zeta}r_{t}^{e}+\frac{\left[  \beta
(1+\bar{r}_{GHH})-1\right]  }{\zeta}E_{t}\hat{\delta}_{t+1},\nonumber\\
\pi_{t}  &  =(1+\bar{r})^{-1}E_{t}\pi_{t+1}+\kappa_{GHH}y_{t}, \label{NKPC GHH}%
\end{align}
where $\kappa_{GHH}=\left(  1-\alpha\right)  \left[  \alpha^{-1}-(1+\bar{r}%
)^{-1}\right]  (1-\theta+\theta/\sigma)^{-1}\left(  \frac{\varepsilon}{\sigma
}-1\right)  $ and $\zeta=\frac{1}{1-\sigma/\varepsilon\left(  1-1/\theta
\right)  }>1$. The relation between the steady state real interest rate
and the debt-to-GDP ratio changes and it is now equal to:

\begin{equation}
\bar{r}_{GHH}=\frac{1}{\beta}\left[  \frac{q\left[  \varepsilon\theta
-\sigma\left(  \theta-1\right)  \right]  }{q\left[  \varepsilon\theta
-\sigma\left(  \theta-1\right)  \right]  -\varepsilon\theta\left(  1-q\right)
\left(  1/\beta-q\right)  \frac{B}{Y}}\right]  -1, \label{ssrbar_GHH}
\end{equation}
while the multiplier for debt becomes:
\begin{equation}
BM_{GHH}=\frac{\left[  1-(1+\bar{r}_{GHH})^{-1}\mu\right]  \left[
\beta(1+\bar{r}_{GHH})-1\right]  }{\frac{\zeta}{\theta}\left[  \beta(1+\bar
{r}_{GHH})\left(  1+\frac{\theta}{\zeta}\phi_{y}\right)  -\mu\right]  \left[
1-(1+\bar{r}_{GHH})^{-1}\mu\right]  +\beta(1+\bar{r}_{GHH})\kappa_{GHH}\left(
\phi_{\pi}-\mu\right)  }. \label{bm olg GHH}%
\end{equation}
Also in this case, we can prove numerically that the size of the multiplier
is lower than one for sensible calibrations of the parameters.

Figure \ref{fig:multiq} and \ref{fig:multi2} show that the multiplier under GHH preferences is higher than under the benchmark log-utility preferences under our benchmark calibration in Table \ref{tab:calibration}. This is due to complementarity between hours and consumption, which is higher under GHH preferences, as there is no wealth effect. Under benchmark preferences, the wealth effect due to the increase in the debt induces households to consume more and work less: the labor supply curve shifts inwards, and therefore mitigates the effects of the fiscal expansion. If households have GHH preferences, the wealth effect on labor supply is absent, thus the labor supply schedule is unaffected, leading to larger increases in hours, consumption and output.\footnote{This mechanism is described in detail in
\citet{monapero08} and \citet{bilbiie11} for changes in government spending in a model with positive interest rates. \citet{cer11jpe} also find this result.}

\subsection{Zero interest rates\label{sec:analitic_multi_ZLB}}

We now turn the attention to the case where the zero lower bound constraint is binding. This occurs for big enough realizations of the shock $r_{t}^{e}$, precisely when $r_{t}^{e}<-\left(  \phi_{\pi}\pi_{t}+\phi_{y}y_{t}\right)$, which implies:
\begin{equation}
i_{t}=0\text{.} \label{zero i}
\end{equation}

We make the same fiscal experiments presented under positive interest rates.
For the benchmark model, the multiplier of government debt is modified as
follows:%
\begin{equation}
BM^{ZLB}=\frac{\left[  \beta(1+\bar{r})-1\right]  \left[  1-(1+\bar
{r})^{-1}\mu\right]  }{\left[  \beta(1+\bar{r})-\mu\right]  \left[
1-(1+\bar{r})^{-1}\mu\right]  -\beta(1+\bar{r})\kappa\mu} \label{bm olg ZLB}%
\end{equation}
The multiplier is positive if:%
\[
\left[  \beta(1+\bar{r})-\mu\right]  \left[  1-(1+\bar{r})^{-1}\mu\right]
-\beta(1+\bar{r})\kappa\mu>0
\]

Note that this condition crucially depends on the persistence of the shock
$\mu$.

Under GHH preferences, we get the following zero lower bound multiplier:
\begin{equation}
BM_{GHH}^{ZLB}=\frac{\left[  1-(1+\bar{r}_{GHH})^{-1}\mu\right]  \left[
\beta(1+\bar{r}_{GHH})-1\right]  }{\left[  1-(1+\bar{r}_{GHH})^{-1}\mu\right]
\left[  \beta(1+\bar{r}_{GHH})-\mu\right]  \frac{\zeta}{\theta}-\mu
\beta(1+\bar{r}_{GHH})\kappa_{GHH}} \label{bm olg ZLB GHH}%
\end{equation}

The literature robustly finds that fiscal multipliers are higher under the ZLB than under positive interest rates. Our results make no exception, because this also holds for the debt multipliers in the OLG context for both specifications of preferences. While Figure \ref{fig:multi2} shows it numerically for our benchmark calibration, it is easy to show it analytically. Under benchmark preferences, the condition for the debt multiplier to be higher under the ZLB than when nominal rates are positive is:
\begin{equation}%
\begin{split}
\left[  \beta(1+\bar{r})\left(  1+\phi_{y}\right)  -\mu\right]  &\left[
1-(1+\bar{r})^{-1}\mu\right]  +\beta(1+\bar{r})\kappa\left(  \phi_{\pi}-\mu\right)
\\
&> \left[  \beta(1+\bar{r})-\mu\right]  \left[  1-(1+\bar{r})^{-1}\mu\right]
-\beta(1+\bar{r})\kappa\mu,
\end{split}
\end{equation}
which always holds as $\phi_{y}\geqslant0$ and $\phi_{\pi}\geqslant\mu$.
Similarly, with GHH preferences the corresponding condition is:
\begin{equation}
\begin{split}
&\left[  1-(1+\bar{r}_{GHH})^{-1}\mu\right]  \left[  \beta(1+\bar{r}_{GHH}%
)-\mu\right]  \frac{\zeta}{\theta}-\mu\beta(1+\bar{r}_{GHH})\kappa_{GHH}
\\
&\quad<\frac{\zeta}{\theta}\left[  \beta(1+\bar{r}_{GHH})\left(  1+\frac{\theta
}{\zeta}\phi_{y}\right)  -\mu\right]  \left[  1-(1+\bar{r}_{GHH})^{-1}%
\mu\right]  +\beta(1+\bar{r}_{GHH})\kappa_{GHH}\left(  \phi_{\pi}-\mu\right)  ,
\end{split}
\label{c1}%
\end{equation}
which is always satisfied as $\phi_{y}\geqslant0$, $\theta,\zeta>1$ and
$\phi_{\pi}\geqslant\mu$.

Intuitively, monetary policy is not reacting to inflation under the ZLB, so
that there is no partial crowding out of the initial increase in aggregate
demand. As shown in \citet{egge11nber}, when nominal interest rates are zero
the AD\ curve shifts from downward sloping, as it is standard in macroeconomic
textbooks, to upward sloping, because the central bank does not offset
inflationary pressures with interest rates hikes. In this case, a higher
inflation means a lower real interest rate, thus implying higher demand. This
is the reason why fiscal multipliers are bigger under the ZLB constraint.

\subsection{Determinacy properties and the multipliers}

The determinacy properties of the equilibrium for both models considered is
crucially dependent on parameter $\mu$, i.e. on the probability of remaining
out of the steady state after the shock. In general, the determinacy
conditions, under both positive and zero interest rates and under both types
of preferences, correspond to the conditions for the multipliers to be
positive. It is important to note that, for both types of preferences under
positive interest rates, the determinacy conditions are always satisfied if
the Taylor principle holds. These conditions for the benchmark model and GHH
preferences are respectively:%
\begin{equation}
\mu^{2}+\beta(1+\bar{r})^{2}\left(  1+\phi_{y}+\kappa\phi_{\pi}\right)  -(1+\bar
{r})\left[  \beta\left(  1+\phi_{y}+\kappa(1+\bar{r})\right)  +1\right]  \mu>0,
\label{det i>0}%
\end{equation}%
\begin{equation}
\mu^{2}+\frac{\beta(1+\bar{r}_{GHH})^{2}\theta}{\zeta}\left\{  \frac{\zeta
}{\theta}+\phi_{y}+\kappa\phi_{\pi}\right\}  -(1+\bar{r}_{GHH})\left\{  \frac
{\beta\theta}{\zeta}\left[  \frac{\zeta}{\theta}+\phi_{y}+(1+\bar{r}%
_{GHH})\kappa_{GHH}\right]  +1\right\}  \mu>0. \label{det i>0 GHH}%
\end{equation}
After some algebra, it can be shown that the denominators of $BM$ and
$BM_{GHH}$ are positive if conditions (\ref{det i>0}) and
(\ref{det i>0 GHH}) are verified, respectively. And, given that the numerators
of $BM$ and $BM_{GHH}$ are always positive, this corresponds also 
to have positive multipliers for debt, thus implying that a debt increase
leads to an output increase (and vice versa).

Interestingly, under zero interest rates, the determinacy conditions
correspond again to the conditions for the multipliers to be positive, but in
this case they are not satisfied for all values of $\mu$. The determinacy
conditions in this case for the benchmark model and GHH preferences are
respectively:%
\begin{equation}
\mu^{2}-(1+\bar{r})\left\{  \beta\left[  1+\kappa(1+\bar{r})\right]  +1\right\}
\mu+\beta(1+\bar{r})^{2}>0 \label{det i=0}%
\end{equation}%
\begin{equation}
\mu^{2}-(1+\bar{r}_{GHH})\left\{  \frac{\beta\theta}{\zeta}\left[  \frac
{\zeta}{\theta}+(1+\bar{r}_{GHH})\kappa_{GHH}\right]  +1\right\}  \mu+\beta
(1+\bar{r}_{GHH})^{2}>0. \label{det i=0 GHH}%
\end{equation}
Given a standard calibration (see Table \ref{tab:calibration}), under the
benchmark model, the condition of determinacy is satisfied if $\mu\leqslant0.70$. For
these values of $\mu$ the positive relationship between the fiscal variables
and output is confirmed. This condition on $\mu$ is even tighter under GHH
preferences: for the same calibration, in this case $\mu$ should be $\leqslant0.58$
for the system to be determined. Again, this condition also establishes that
the multipliers are positive.


\subsection{Government spending multipliers}
Despite not being the main focus of the paper, we briefly report here the government spending multipliers. These multipliers regard a change in government spending at balance budget, that is, coupled with an one-off change in taxation and holding the public debt constant.
For the benchmark case, the government spending multiplier is:\footnote{The multiplier is computed as $dy^{S}/dg^{S}$. Remember that $g_{t}=\frac{G_{t}}{Y}$ as the steady state of $G_{t}$ is assumed to be zero. Thus, a $1\%$ increase in spending with respect to steady state output leads to a $GSM\%$ increase in output.}
\begin{equation}
    GSM=\frac{\left[  \beta(1+\bar{r})-\mu\right]  \left[  1-(1+\bar
{r})^{-1}\mu\right]  +\beta(1+\bar{r})\kappa\left(  1-L\right)  \sigma\left(
\phi_{\pi}-\mu\right)  }{\left[  \beta(1+\bar{r})\left(  1+\phi_{y}\right)
-\mu\right]  \left[  1-(1+\bar{r})^{-1}\mu\right]  +\beta(1+\bar{r})\kappa\left(
\phi_{\pi}-\mu\right)  
}.
\label{GSM}
\end{equation}

The numerator in this expression is higher than the one in (\ref{bm olg}), so that the multiplier of spending is higher than the one of debt. This result is in line with recent literature on fiscal multipliers for
spending and taxes \citep[e.g.,][]{egge11nber,deve11,cer11jpe}. However, these are very different fiscal policy measures so the comparison is not really meaningful. In the case of government spending the wealth effect is negative because agents will have to pay higher taxes.\footnote{We calculate the multiplier using the system (\ref{AD})-(\ref{delta}), so that means that taxes adjust endogenously to the change in government spending to keep $b'$ constant. Hence, there is no effect coming from postponement of taxes and failure of Ricardian equivalence. In this respect the multiplier should be the same as in an ILRA model. However, the  multiplier is larger in our OLG framework than in an ILRA one, because the gross real interest rate is higher than $1/\beta$, as explained in Section \ref{benchmark}.} Private consumption will contract, but less than the increase in government spending, so that aggregate demand increases.
The presence of price rigidities implies also a shift in the labor demand because some firms will not be able to adjust prices and thus need to increase production to meet the higher demand \citep[see][]{monapero08}. Moreover, the negative wealth effect leads to  an increase in labour supply.

In the GHH case, the government spending multiplier is:
\begin{equation}
GSM_{GHH}=\frac{\left[  1-(1+\bar{r}_{GHH})^{-1}\mu\right]  \left[  \beta(1+\bar
{r}_{GHH})-\mu\right]  \zeta}{\frac{\zeta}{\theta}\left[  \beta(1+\bar
{r}_{GHH})\left(  1+\frac{\theta}{\zeta}\phi_{y}\right)  -\mu\right]  \left[
1-(1+\bar{r}_{GHH})^{-1}\mu\right]  +\beta(1+\bar{r}_{GHH})\kappa_{GHH}\left(
\phi_{\pi}-\mu\right),}
\label{GSMGHH}
\end{equation}
which is larger than the debt multiplier, as $\zeta>1$ and $\mu<1$.

Regarding the ZLB, in accordance with the literature, our framework yields larger government spending multiplier under the ZLB than under positive interest rates for both specification of preferences. These are equal to for the benchmark case and the GHH case, respectively:
\begin{equation}
GSM^{ZLB}=\frac{\left[ \beta (1+\bar{r})-\mu \right] \left[ 1-(1+\bar{r})^{-1}\mu \right] -\beta (1+\bar{r})k\left( 1-L\right) \sigma \mu }{\left[\beta (1+\bar{r})-\mu \right] \left[ 1-(1+\bar{r})^{-1}\mu \right] -\beta (1+\bar{r})k\mu} 
\end{equation}

\begin{equation}
GSM_{GHH}^{ZLB}=\frac{\left[
1-(1+\bar{r}_{GHH})^{-1}\mu\right]  \left[  \beta(1+\bar{r}_{GHH})-\mu\right]
\zeta}{\frac{\zeta}{\theta}\left[  1-(1+\bar{r}_{GHH})^{-1}\mu\right]  \left[  \beta(1+\bar
{r}_{GHH})-\mu\right]  -\mu\beta(1+\bar{r}_{GHH})\kappa_{GHH}}
\end{equation}
Finally, we can prove analytically that the multiplier of debt is lower than the multiplier of spending for the GHH preferences model (as $\mu<1$ and $\zeta>1$), while for the benchmark model, this holds numerically for the standard calibration.

\FloatBarrier

\section{The size of the fiscal multipliers and the cost of fiscal consolidation\label{sec:mult_numerical}}

In the previous section, we provided analytical results on debt (and government spending) multipliers in our OLG framework. In this section we simulate the model numerically, both under positive interest rates and at the ZLB. Our aim is twofold. First, we want to quantify the size of the multiplier associated to a cut in lump-sum taxes that raises the level of debt. 
Second, we want to shed light on the effects of this type of intervention during a liquidity trap, with focus on the role of the initial level of debt. Moreover, \cite{lindetraba18} show that fiscal multipliers computed using the non-linear solution of a DSGE model can be much smaller than the multipliers associated with the linear solution, especially in a ZLB episode where non-linearity matters most. Hence, to check that our analytical results in the linearized model still hold qualitatively in the non-linear case, we simulate our OLG model in its original non-linear form.

The calibration of the structural parameters is reported in Table \ref{tab:calibration}. We adopt standard quarterly values, following the recent literature in this field. To pin down the labor disutility parameter $\eta$, we assume that people spend at work roughly one third of their time endowment, so that hours worked are set to 0.3 at steady state. This calibration ensures that the non-stochastic steady state level of output is unique and independent of the debt size. The elasticity of substitution between goods is 6, corresponding to a 20\% mark up over marginal costs. We assume constant returns to scale, i.e. $\sigma$ equal to 1. The Calvo probability $\alpha$ of keeping prices fixed is set to 0.75, implying an average duration of prices of one year. For monetary policy, we assume a zero inflation target and set $\phi_{\pi}=2$ so that the Taylor principle holds. The response of nominal interest rates to output $\phi_{y}$ is set to 0.125.
To calibrate the subjective discount factor ($\beta$) and the survival probability ($q$) we exploit the relationship between the debt level and the real interest rate at steady state. Figure \ref{fig:rr_scatter} illustrates this relationship in the data, while equations \eqref{ssrbar} and \eqref{ssrbar_GHH} describe it in the model. Hence, we choose the values of $\beta$ and $q$ so the relations \eqref{ssrbar} and \eqref{ssrbar_GHH}  closely match the empirical counterpart.
\footnote{More precisely, we fix $\beta$ and $q$ so that the interest rates implied by \eqref{ssrbar} and \eqref{ssrbar_GHH} are identical to the fitted values in Figure \ref{fig:rr_scatter} when the debt-to-GDP level is equal to 60\% and 100\%, respectively. Note that the relationship between $\bar{r}$ and $B/Y$ in (\ref{ssrbar}) and \eqref{ssrbar_GHH} is almost linear in the range that we consider.} As a result, $\beta$ is equal to 0.998 for both types of preferences, while $q$ is equal to 0.9512 under log-log preferences and to  0.9785 under GHH preferences. 
This implies different time horizons for households in their planning decisions under the different types of preferences, about five and twelve years, respectively.
Note that these values are considerably higher than the one usually assumed in the OLG literature.\footnote{\citet{bayosghe06} estimate values for $q$ between 0.8 and 0.88. \citet{CasteNisti10} estimated it to be between 0.82 and 0.92, with a posterior mode of 0.87. These values are below the values we are considering here and more generally way below what is generally assumed in models where $q$ is calibrated according to demographics.} We are therefore quite conservative in our choice of $q$, pushing the model towards the ILRA case and the delivering of small debt multiplier. We will briefly investigate how multipliers change with $q$ below. We initially set the debt-to-GDP ratio to 60\% on an annual basis (i.e., 2.4 on a quarterly basis) in line with the reference value for the euro area. 
In Section \ref{sec:multi_BY} we will discuss the implications of a higher debt burden.

\begin{table}[h] \centering
\refstepcounter{table} \label{tab:calibration}
\small{
\begin{tabular}{ccl}
\multicolumn{3}{l}{Table \ref{tab:calibration}. Parameters calibration} \\
\noalign{\smallskip}\toprule\noalign{\smallskip}  
\textbf{Parameter} & \multicolumn{1}{c}{\textbf{Value}} & \textbf{Description}\\
\noalign{\smallskip}\hline\noalign{\smallskip}  
$\beta$         & 0.998 &  Intertemporal discount factor\\[0.5ex]
\multirow{2}{*}{$q$}             & 0.9512 &    Survival probability, log-log preferences\\
                & 0.9785 &    Survival probability GHH preferences\\[0.5ex]
$\theta$        & 6 &      Elasticity of substitution between goods\\[0.5ex]
$\varepsilon$   & 2 &       Inverse of labor supply elasticity (GHH preferences only)\\[0.5ex]
$\alpha$        & 0.75 &    Calvo price stickiness\\[0.5ex]
$\sigma$        & 1 &       Output elasticity of labor\\[0.5ex]
$\phi_{\pi}$    & 2 &       Taylor rule reaction to inflation\\[0.5ex]
$\phi_{y}$      & 0.125 &   Taylor rule reaction to output\\[0.5ex]
$B/Y$           & 2.4 &     Debt-to-GDP ratio (same as 60\% in annualized terms)\\
\noalign{\smallskip}\bottomrule  
\end{tabular}%
}
\end{table}%

The simulation exercise is similar to the one we studied in the analytical part of the paper and is based on two main elements: a preference shock that mimics an economic downturn and a fiscal stimulus given by an increase in debt (i.e., a tax cut) at the outset of the recession. 
More precisely, we start from the steady state and assume that a persistent shock to the intertemporal discount factor 
hits the economy and brings it in a recession that lasts for two years (i.e., 8 periods in our quarterly calibration). 
The government responds immediately to the adverse shock with an increase in the level of debt, engineered by a one-period tax cut.
The increase in debt can be either temporary or permanent. In the first case, debt is kept constant at the new plateau throughout the recession, until a tax hike in period 9 restores the initial debt level. In the second, debt is permanently fixed at the new level.\footnote{If the change in debt is permanent, the economy moves to another steady state once the preference shock is reabsorbed. The new steady state will have a higher debt-to-GDP ratio and, given equations \eqref{ssrbar} and \eqref{ssrbar_GHH}, a higher real interest rate. Instead, the level of output is determined by labor supply and remains unaltered.}
Note that in both cases, after the initial period, the fiscal plan is assumed to be credible and perfectly anticipated by private agents. 
In addition, by varying the size of the preference shock we are able to reproduce both a mild recession scenario, in which interest rates remain above zero, and a severe recession scenario, in which the ZLB is activated. 
Moreover, we consider both benchmark and GHH preferences.
In each exercise, we first run a baseline simulation of the model with the preference shock only, and then we run a second simulation adding the debt shock. By subtracting the time series generated by the two simulations, we are able to obtain the impulse responses following a debt shock and the associated debt multiplier.\footnote{We use Dynare version 4.5.7  \citep[see][]{Dynare2011} to run deterministic simulations of the model, in which perfect foresight is assumed.}

\begin{figure}[h!]
\begin{center}
    \includegraphics[trim = 20mm  20mm 20mm  20mm, clip, scale = 0.65]{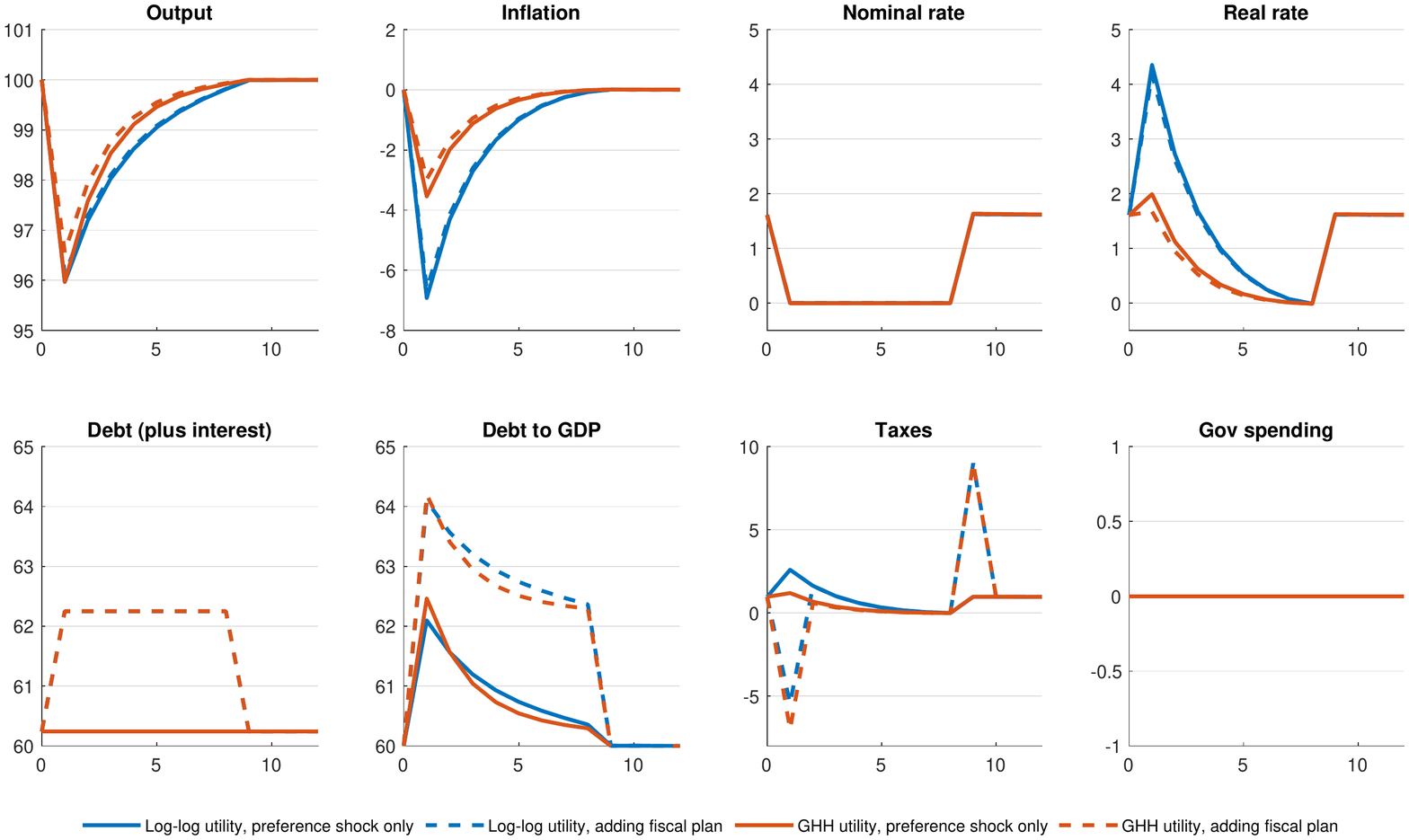}
    \caption{Impulse responses to a preference shock with and without fiscal intervention. \newline {\protect\footnotesize \textit{Notes:}  Inflation and interest rates are expressed in annualized percentage points. Taxes are expressed as percentage of quarterly output.  Debt is expressed as percentage of annual output. } }
    \label{fig:risposta_bperm_sovrapposte}
\end{center}
\end{figure}

Figure \ref{fig:risposta_bperm_sovrapposte} shows the outcome of the simulations for the case in which the preference shock is strong enough to bring the economy to the ZLB.\footnote{We choose the size of the preference shock so that output falls on impact by 4\%. A similar figure can be reproduced for the situation where the economy enters a mild recession and remains above the ZLB.} 
Let us first consider the solid blue lines, which correspond to the baseline simulation under log-log preferences. In period 1 an adverse preference shock hits the economy and generates a severe recession: output falls by 4\%, inflation also drops and the central bank reacts by bringing the nominal rate to its zero limit. 
Note that the government avoids any fiscal intervention and maintains real debt (plus interest) at its steady state level throughout the recession. 
To do so, taxes are adjusted upwards to meet the higher debt service costs brought about by the increase in real rates. 
The debt-to-GDP ratio also rises temporarily until output fully recovers. 
Now consider the effect of the fiscal stimulus.
The dashed blue lines correspond to the simulation that incorporates a temporary increase in debt of 2\% of annual output (or 8\% of quarterly output), engineered with a pure postponement in tax payments: the initial tax cut is coupled with a single tax hike at the end of the recession. The fiscal plan induces a positive wealth effect on consumption that leads to an increase in output and inflation with respect to the baseline simulations. These effects are however quite small under benchmark preferences.
The red lines illustrate the outcome of the same simulations when we assume GHH preferences. As discussed above, this specification of utility eliminates the disincentive to supply labor when consumption increases, thus reinforcing the expansionary effect of the fiscal plan. As a result, output falls on impact by less than 3.5\%, which means that it remains more that 0.5\% above the path it would have followed had debt been kept constant. 

\begin{table}[h] \centering
\refstepcounter{table} \label{tab:simu_multipliers}

\begin{tabularx}{.85\textwidth}{XD{.}{.}{3.5}D{.}{.}{3.5}cD{.}{.}{3.5}D{.}{.}{3.5}}
\multicolumn{6}{l}{Table \ref{tab:simu_multipliers}. Simulated debt multipliers for an expansion in debt} \\
\noalign{\smallskip}\toprule\noalign{\medskip} 
\multicolumn{6}{l}{\emph{Panel a. Multipliers under log-log preferences}} \\[2ex]
& \multicolumn{2}{c}{Temporary change} &\qquad &\multicolumn{2}{c}{Permanent change}\\
\multicolumn{1}{l}{}& \multicolumn{1}{c}{Normal times} & \multicolumn{1}{c}{ZLB} && \multicolumn{1}{c}{Normal times} & \multicolumn{1}{c}{ZLB}\\
\cline{1-6}\noalign{\smallskip}     
\multicolumn{6}{l}{\emph{Impact }} \\[0.5ex]  
\multicolumn{1}{l}{$q=0.9512$}	& 0.0006 &  0.0189  &&  0.0000 &  0.0387  \\[0.5ex]
\multicolumn{1}{l}{$q=0.95$}	& 0.0006 &  0.0199  &&  0.0000 &  0.0407 \\
\multicolumn{1}{l}{$q=0.96$}	& 0.0004 &  0.0128  &&  0.0000 &  0.0261 \\
\multicolumn{1}{l}{$q=0.97$}	& 0.0002 &  0.0072  &&  0.0000 &  0.0148 \\
\multicolumn{1}{l}{$q=0.98$}	& 0.0001 &  0.0033  &&  0.0000 &  0.0067  \\
\multicolumn{1}{l}{$q=0.99$}	& 0.0000 &  0.0009  &&  0.0000 &  0.0018 \\[2ex]
\multicolumn{6}{l}{\emph{Present value}} \\[0.5ex]  
\multicolumn{1}{l}{$q=0.9512$}	& 0.0010 &  0.0073  &&  0.0000 &  0.0131  \\[0.5ex]
\multicolumn{1}{l}{$q=0.95$}	& 0.0011 &  0.0076  &&  0.0000 &  0.0138 \\
\multicolumn{1}{l}{$q=0.96$}	& 0.0007 &  0.0049  &&  0.0000 &  0.0088 \\
\multicolumn{1}{l}{$q=0.97$}	& 0.0004 &  0.0028  &&  0.0000 &  0.0050 \\
\multicolumn{1}{l}{$q=0.98$}	& 0.0002 &  0.0013  &&  0.0000 &  0.0023 \\
\multicolumn{1}{l}{$q=0.99$}	& 0.0000 &  0.0003  &&  0.0000 &  0.0006 \\
\noalign{\smallskip} \midrule\noalign{\medskip} 
\multicolumn{6}{l}{\emph{Panel b. Multipliers under GHH preferences}} \\[2ex]
& \multicolumn{2}{c}{Temporary change} &\qquad &\multicolumn{2}{c}{Permanent change}\\
\multicolumn{1}{l}{}& \multicolumn{1}{c}{Normal times} & \multicolumn{1}{c}{ZLB} && \multicolumn{1}{c}{Normal times} & \multicolumn{1}{c}{ZLB}\\
\cline{1-6}\noalign{\smallskip}     
\multicolumn{6}{l}{\emph{Impact }} \\[0.5ex]  
\multicolumn{1}{l}{$q=0.9785$}	& 0.0004 &  0.0707  &&  0.0000 &  0.1387 \\[0.5ex]
\multicolumn{1}{l}{$q=0.95$}	& 0.0022 &  0.3766  &&  0.0006 &  0.7439 \\
\multicolumn{1}{l}{$q=0.96$}	& 0.0013 &  0.2380  &&  0.0003 &  0.4686  \\
\multicolumn{1}{l}{$q=0.97$}	& 0.0007 &  0.1359  &&  0.0001 &  0.2672 \\
\multicolumn{1}{l}{$q=0.98$}	& 0.0003 &  0.0614  &&  0.0000 &  0.1205 \\
\multicolumn{1}{l}{$q=0.99$}	& 0.0001 &  0.0167  &&  0.0000 &  0.0328  \\[2ex]
\multicolumn{6}{l}{\emph{Present value}} \\[0.5ex]  
\multicolumn{1}{l}{$q=0.9785$}	& 0.0013 &  0.0230  &&  0.0000 &  0.0401 \\[0.5ex]
\multicolumn{1}{l}{$q=0.95$}	& 0.0068 &  0.1257  &&  0.0006 &  0.2216 \\
\multicolumn{1}{l}{$q=0.96$}	& 0.0044 &  0.0785  &&  0.0003 &  0.1378 \\
\multicolumn{1}{l}{$q=0.97$}	& 0.0025 &  0.0444  &&  0.0001 &  0.0777 \\
\multicolumn{1}{l}{$q=0.98$}	& 0.0011 &  0.0200  &&  0.0000 &  0.0348 \\
\multicolumn{1}{l}{$q=0.99$}	& 0.0003 &  0.0054  &&  0.0000 &  0.0094 \\
\noalign{\smallskip} \bottomrule
\end{tabularx}
{\footnotesize \begin{tabularx}{0.85\textwidth}{X}
{\footnotesize {\bf Notes:}
Impact multipliers are computed as the ratio between the impact variation of output (induced by the fiscal shock only) and the impact variation of debt.
In the temporary debt plan, debt is raised in period 1 and brought back to the initial level in period 9, when the recession ends.
In the permanent debt plan, debt is raised once and for all in period 1. 
Present value multipliers are computed using the formula in footnote \ref{footnote:PV}, setting $k=8$.
}
\end{tabularx}}
\end{table}

The debt multipliers resulting from the simulations are shown in Table \ref{tab:simu_multipliers}. To assess the overall effect of the fiscal plan, we consider both the impact multiplier and the present value multiplier, where the latter is computed as the present value of the variation in output during the recession period divided by the present value of the variation in debt.\footnote{\label{footnote:PV}Following \citet{MountfordUhlig2009}, present value multipliers are obtained with the formula:
$$\text{Present value multiplier }(k) = \frac{\sum_{j=0}^k\prod_{i=0}^j(1+r_{t+i})^{-1}\Delta Y_{t+j}}{\sum_{j=0}^k\prod_{i=0}^j(1+r_{t+i})^{-1}\Delta B'_{t+j}}.$$
Note the present value multipliers coincides with the impact multiplier when $k=0$.}
Under both types of preferences and both types of fiscal plans, debt multipliers are negligible in a mild recession scenario, but they are of one or two orders of magnitude larger when a severe recession hits the economy and monetary policy is constrained by the ZLB. In this case, the central bank does not adjust nominal rates upwards when the government implements the stimulus plan, as even lower rates would be required to contrast the adverse preference shock. 
As explained in \citet{egge11nber}, at the ZLB the aggregate demand curve is actually upward sloping---rather than downward sloping as in standard theory---because the central bank does not offset inflationary pressures with interest rates hikes.
In this case, a higher inflation means a lower real interest rate, implying higher demand and, in turn, larger fiscal multipliers.

The effectiveness of fiscal interventions depends on the reaction of monetary policy: when this reaction is muted, debt multipliers are more sizeable. But are they large enough to convince us that the debt level is a relevant policy instrument during a liquidity trap? Our simulations indicate that this can indeed be the case, but three elements are required. Two of them pertain to the structure of the model and are given by the size of the survival probability and the form of the utility function. 
When $q$ gets smaller, the model becomes increasingly less Ricardian and the wealth effects of debt play a more significant role, so that multipliers are magnified. Moreover, when households have GHH preferences rather than log-log preferences, the increase in consumption does not depress labor supply, and this again contributes to making multipliers bigger. The third element is instead related to the design of the fiscal plan. A permanent increase in debt produces a stronger response in output, as agents anticipate that the initial tax cut will not be reverted at the end of the recession. If we combine these three elements---i.e., set $q$ to 0.95, choose GHH preferences, and consider a permanent expansion in debt---we obtain a debt multiplier equal to 0.74 on impact, a figure that is not so distant from the estimates of the government spending multipliers reported by the recent literature. If we consider only a temporary expansion in debt, the associated impact multiplier reduces to 0.38, which may be considered as a small yet not irrelevant value. 


Note that we assume that the fiscal stimulus only alleviates the fall in output caused by the preference shock, as the tax cut is too small to bring the economy back into the positive interest rates territory: the ZLB episode ends only when the preference shock eventually disappears, as shown in Figure \ref{fig:risposta_bperm_sovrapposte}.
\footnote{We also tried experimenting with larger fiscal shocks that push the economy out of the ZLB before the end of the recession. In this case the multipliers lie about halfway between the values obtained under normal times and at the ZLB, reported in Table \ref{tab:simu_multipliers}. Results are available from the authors upon request.} 

Table \ref{tab:simu_multipliers} also implicitly illustrates the cost of a fiscal contraction engineered through the opposite policy, that is, through a temporary decrease in public debt caused by an initial tax hike followed by a cut.\footnote{Since the model is non-linear, the multipliers associated temporary decrease in public debt may
differ from those reported in Table \ref{tab:simu_multipliers}, which correspond to an increase in debt. Nonetheless, our simulations indicate that multipliers are only slightly asymmetric, so that Table \ref{tab:simu_multipliers} remains valid for describing the effect of a fiscal contraction.} The main message is that fiscal consolidation through a temporary decrease in government debt may not be too costly in terms of output losses. However, this is true only as long as there are wealth effects on the labor supply (as in the case of log-log preferences) 
and the ZLB is not binding. This consideration emphasizes once again the perils of implementing a fiscal consolidation during a severe recession, when the central bank is unable to mitigate the negative consequences on output and prices. A fiscal restriction has an opposite effect on demand, and this effect will be amplified by the mechanism explained above. A restrictive fiscal intervention has deflationary effects, and when interest rates are constrained at the zero level, the central bank could not cut the interest rates to counteract deflation. 
Thus, the real interest rate will be higher leading to even lower demand.

\FloatBarrier

\subsection{Debt multipliers and the level of debt\qquad\label{sec:multi_BY}}

In the previous section we discussed the size of debt multipliers implied by our model when the stock of public debt amounts to 60\% of GDP. A natural question is how previous findings change once we consider a different level of debt, and, in particular, once we increase the steady state debt-to-GDP ratio to match the values recently reached by most industrialized countries after the 2008 recession.

The analytic results of Sections \ref{sec:analitic_multi_noZLB} and \ref{sec:analitic_multi_ZLB} showed that the multipliers of the linearized version of the model directly depend on the steady state real interest rate, both in normal times and during a ZLB episode. In turn, the steady state real rate depends on the steady state debt-to-GDP ratio. Given our calibration strategy, Figure \ref{fig:rr_scatter} also illustrates the relations (\ref{ssrbar}) and (\ref{ssrbar_GHH}). The implied increase in rates is pronounced: when the debt-to-GDP\ ratio passes from 60\% to 200\%, the annualized steady state real rate rises from 1.62\% to 3.65\%. 
As such, the level of debt is potentially a key variable for the size of debt multipliers.



\begin{figure} [ht]
    \centering
   \includegraphics[trim=0mm 60mm 0mm 60mm, clip,scale = 0.60]{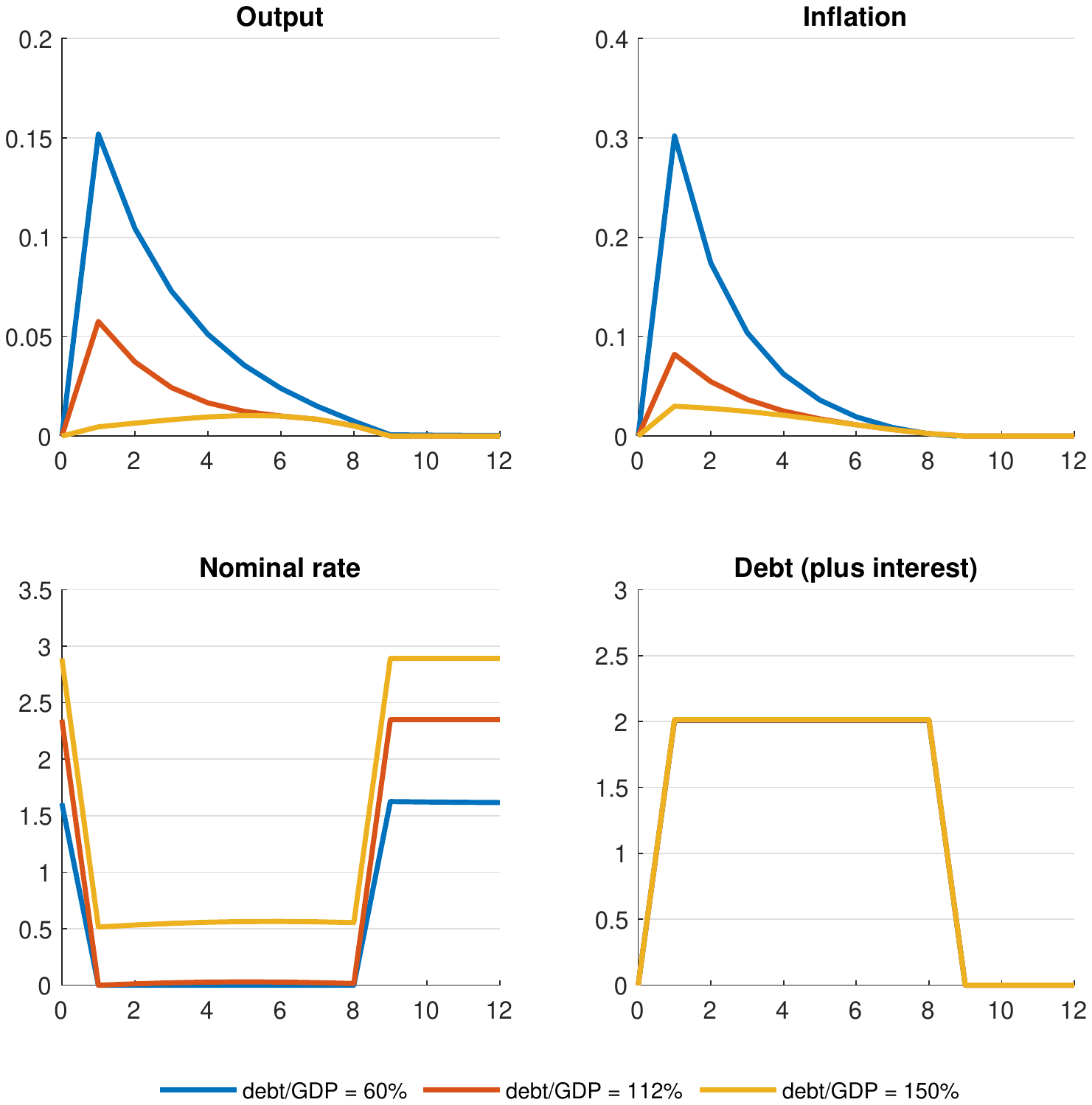}
   \caption{Impulse responses to a temporary increase in debt. \newline {\protect\footnotesize \textit{Notes:} 
   Impulse responses for output, inflation and debt-to-GDP represent the difference between the series generated with both the preference shock and the fiscal plan and the series generated with the preference shock only. The fiscal stimulus is given by a tax cut that raises the debt-to-GDP ratio by 2 percentage points in period 1, followed by a tax hike in period 9 that brings debt to its steady state level.    
   All variables are in percentage points. The nominal interest rate and inflation are annualized. In the bottom-right panel, we show the ratio between the variation in debt and the steady state level of annual output. 
   } }
   \label{fig:impulse}
\end{figure}

To further examine the relation between the debt multiplier and the level of debt, we run another set of simulations in which we fix the size of the preference shock and we evaluate the effect of an expansion in debt for different levels of the initial debt-to-GDP ratio. Results are reported in Figure \ref{fig:impulse}, where we show impulse responses of the economy after a temporary debt plan implemented at the ZLB, with log-log preferences. The preference shock is calibrated so that output contracts by 4\% when the debt-to-GDP ratio is 60\%. In this case the nominal rate remains at the ZLB for the entire duration of the recession. In the bottom-right panel of Figure \ref{fig:impulse}, we show the ratio between the variation in debt and the steady state level of annual output, which is the same for all the cases considered.\footnote{More precisely, the blue lines correspond to a temporary increase of the annualized debt-to-GDP ratio from 60\% to 62\%, the red lines to an increase from 112\% to 114\%, the yellow lines to an increase from 150\% to 152\%.}
For a 60\% debt-to-GDP ratio, the figure implies a debt multiplier of magnitude 0.0189 (as in Table \ref{tab:simu_multipliers}), computed as the ratio between the impact increase of output (0.151\%, in terms of quarterly output units) and the impact variation in debt (8\%, again in quarterly terms).


For a debt-to-GDP ratio of 112\%, the economy still falls into the ZLB on impact, but nominal rates can be lifted above zero already in second period. The intuition is that the steady state real rate increases with the level of steady state debt. It follows that the central bank starts from a higher nominal rate, for a given inflation target, and thus it has additional room to reduce rates and contrast the adverse preference shock. Being out from the ZLB, monetary policy is able to partially offset the positive demand shock represented by fiscal intervention, so that the responses of output and inflation are largely muted if compared to the 60\% debt-to-GDP scenario.  
When debt is equal to 150\% of GDP, the central bank can maintain interest rates above the zero limit and the expansionary effect of the fiscal stimulus plan is even weaker. The implied debt multiplier is negligible.
The main takeaway here is that the steady state level of debt-to-GDP ratio could provide an alternative route to keep the economy away from the ZLB. Some papers in the literature \citep[e.g.,][]{BlanchardDellAricciaMauro2010} suggested to increase the inflation target to increase the steady state nominal rate (for a given steady state real rate) in order to provide monetary policy with more room for manoeuvre in case of a deflationary shock. This generated some debate in the literature \citep[for a recent survey see][]{ascasbo14}.
Here, we uncover an alternative route, based on the fact that in an OLG model (as well as in the data) the steady state real rate is an increasing function of the debt-to-GDP ratio. Hence, a rise in the debt-to-GDP ratio could provide monetary policy with more room for manoeuvre in case of a deflationary shock, because it increases the steady state nominal rate, through a rise in the steady state  \textit{real} rate (for a given steady state \textit{inflation target}).

\FloatBarrier

\section{Conclusions \label{sec:conclusions}}
This paper analyzes the debt multiplier, that is, it addresses the consequences on economic activity of a temporary change in government debt in a simple OLG model, in which agents exhibit non-Ricardian behavior. Variations in lump-sum taxes fully reversed in the future generate wealth effects that influence the consumption and labor choices. One could consider the debt multiplier as a complement to the fiscal multipliers so far appeared in the literature, mainly featuring an ILRA framework. The debt multiplier would be the additional multiplier that one can get if any temporary fiscal policy measure would be financed through a pure temporary increase in public debt (i.e., a postponement of lump-sum taxes). 
We show that the debt multiplier is positive in an OLG framework. As for other fiscal multipliers analyzed in the literature, the debt multiplier depends critically on the stance of monetary policy and it is larger when interest rates are stuck at the zero bound and the subsequent adjustment in taxes is delayed until the recession has ended. Moreover, the debt multiplier increases with the level of debt-to-GDP ratio. Finally, we point out that, since the steady state real interest rate is increasing with the steady state level of debt-to-GDP ratio, a rise in the debt-to-GDP ratio could provide an alternative way to stay away from the ZLB.

\singlespacing
\small
\bibliographystyle{ecta}
\bibliography{fiscal_multi}

\end{document}